\documentclass[]{spie}  

\usepackage{amsmath,amsfonts,amssymb}
\usepackage{graphicx}
\usepackage{caption}
\usepackage{subcaption}
\usepackage[colorlinks=true, allcolors=blue]{hyperref}

\title{On-sky demonstration at Palomar Observatory of the near-IR, high-resolution VIPA spectrometer}

\author[a]{Alexis Carlotti}
\author[a]{Alexis Bidot}
\author[a]{David Mouillet}
\author[a]{Jean-Jacques Correia}
\author[a]{Laurent Jocou}
\author[a]{Stéphane Curaba}
\author[a]{Alain Delboulbé}
\author[a]{Etienne Le Coarer}
\author[a]{Patrick Rabou}
\author[b]{Guillaume Bourdarot}
\author[a]{Thierry Forveille}
\author[a]{Xavier Bonfils}
\author[c,d]{Gautam Vasisht}
\author[c,d]{Dimitri Mawet}
\author[c,d]{Rick S. Burruss}
\author[e]{Rebecca Oppenheimer}
\author[f,g]{René Doyon}
\author[f,g]{Etienne Artigau}
\author[f,g]{Philippe Vallée}
\affil[a]{Univ. Grenoble Alpes, CNRS, IPAG, 38000 Grenoble, France}
\affil[b]{Max Planck Institute for extraterrestrial Physics, 85748 Garching, Deutschland}
\affil[c]{Department of Astronomy, California Institute of Technology, Pasadena, CA, USA}
\affil[d]{Jet Propulsion Laboratory, Pasadena, CA, USA}
\affil[e]{Astrophysics Department, American Museum of Natural History, New York, NY 10024, USA}
\affil[f]{Université de Montréal, Département de Physique, IREX, Montréal, QC H3C 3J7, Canada}
\affil[g]{Observatoire du Mont-Mégantic, Université de Montréal, Montréal, QC H3C 3J7, Canada}

\authorinfo{Further author information: (Send correspondence to Alexis Carlotti)\\A.C.: E-mail: alexis.carlotti@univ-grenoble-alpes.fr}

\begin{document} 
\maketitle

\begin{abstract}
A near-IR high-resolution, R$\approx$80000 spectrometer has been developed at IPAG to directly characterize the atmosphere of exoplanets using adaptive optics (AO) assisted telescopes, and a single-mode fiber-injection unit. A first technical test with the 200' Hale telescope at Palomar Observatory occurred in March 2022 using the PALM3000 AO system offered by this telescope. Observations have also been made at the same time with the PARVI spectrometer so that a direct comparison can be made between the two instruments.\\
This spectrometer uses a virtually imaged phased array (VIPA) instead of an echelle grating, resulting in a very compact optical layout that fits in a 0.25m$^3$ cryostat. Using a quarter of an H2RG detector, the spectrometer analyses the middle part of the H-band, from 1.57 to 1.7 microns for 2 sources whose light is transferred from the telescope to the spectrometer using single-mode fibers. By design, the transmission of the spectrometer is expected to be 40-50\%, which is 2-3 times higher than the transmission of current high-resolution spectrometers such as CRIRES+ and NIRSPEC. A damaged cross-disperser limited it to 21\%, however. A replacement grating with a correct, twice as high efficiency has been procured after the on-sky demonstration.\\
In addition to recalling the main specifications of the VIPA spectrometer, this paper presents the control software, the calibration process, and the reduction pipeline that have been developed for the instrument. It also presents the results of the on-sky technical test with the Hale telescope, as well as measurements of the effective resolution and transmission, along with a comparison of a spectrum of the sun obtained with the spectrometer with the BASS2000 reference spectrum. Planned modifications are also discussed. That includes the integration of a new dedicated H2RG detector, and of K-band optics.
\end{abstract}


\section{Introduction}
\label{sec:intro}  

Spectrally characterizing exoplanets atmosphere will be necessary to answer some of the most pressing questions we have about them:
\begin{itemize}
    \item how do planets form? two main mechanisms have been proposed, namely core accretion and gravitational instability. The carbon to oxygen ratio may help determining which one is the most likely to explain the properties of planets,
    \item how habitable are planets? without the right atmosphere (in terms of physical properties and composition), planets in the so-called habitable zone may not be habitable at all,
    \item is there a biological activity on other planets? 
\end{itemize}

In all three cases, a sufficiently high resolution will be required. A resolving power $R=\frac{\lambda}{\Delta \lambda}$ of a few thousands is enough to detect the presence of some molecules such as $H_2 O$, $CH_4$, and $CO$, but a few tens of thousands is necessary to resolve the molecular lines, from which molecular abundances can be derived. The planet velocity can also be measured by analyzing the broadening of the absorption lines. 


The presence of the host star is an obvious obstacle to the characterization of the planet. Spatially resolving the planet, i.e., imaging it, lessens this issue by greatly decreasing the photon noise of the star. The larger the telescope, and the better the adaptive optics system, the lower the photon noise from unwanted stellar light.

The current high-contrast imaging instruments (VLT-SPHERE\cite{Beuzit2019}, GPI\cite{Macintosh2018}, SCExAO\cite{Ahn2021}, MAGAO-X\cite{Males2018}) only have low-resolution capabilities, however, with $R=50-70$ in the near-IR.

Three projects so far aim at injecting the light of a planet (observed using an extreme adaptive optics system) into a single-mode fiber, and feeding a high-resolution spectrometer with it:
\begin{itemize}
    \item the Keck Planet Imager and Characterizer (KPIC\cite{Mawet2018}), an extreme adaptive optics system that sends light to NIRSPEC\cite{McLean1998} (R=37000, 0.72'' slit width)
    \item HiRISE\cite{Vigan2018,Otten2021}, a fiber-injection unit for VLT-SPHERE, that plans to use CRIRES+\cite{Brucalassi2018SPIE} (R=50000-100000, 0.4-0.2'' slit width )
    \item REACH\cite{Kotani2020}, a fiber-injection unit connecting Subaru-SCExAO with the Infrared Doppler (IRD) instrument\cite{Kotani2018} (R=100000)
\end{itemize}

In all cases, the total transmission is very low. HiRISE expects an average transmission of 3.6\% (from the telescope entrance to the detector of CRIRES+), and KPIC demonstrated a transmission close to 5\%. Multiple attenuation factors exist. Most have already been minimized, for instance by using highly transmissive or reflective optics. One, however, is both problematically low, and can still be significantly increased: the spectrometer's efficiency.

The NIRSPEC and CRIRES+ high-resolution spectrometers have been designed to look at seeing limited sources, and not diffraction limited sources. This limits their transmission. It is for instance 17-22\% for CRIRES+ in the H-band, and an additional loss of 8\% is induced when coupling the fiber coming from SPHERE/HiRISE to CRIRES+. 

It also limits the minimum size of the spectrometers, since the resolution of an échelle spectrograph designed for seeing-limited sources is inversely proportional to the telescope diameter. An instrument like CRIRES+ is already 2m across. This is a potential issue with the 30 to 40m upcoming extremely large telescopes (ELT).

Designing spectrometers for diffraction-limited sources has several advantages\cite{Crass2019}. From an astrophysics point-of-view, it can in particular remove the modal noise that spectrometers fed with multimode fibers suffer from, which is critical for the detection of planets by radial velocity instruments. From a technological point-of-view, it leads to instrument whose size is not set by the telescope diameter, and are intrinsically quite compact, and thus less expensive that their seeing-limited counterparts.

The VIPA spectrometer that has been developed at IPAG since 2017\cite{Bourdarot2017} is designed to look at diffraction-limited sources. It operates with a $\lambda^2$ elementary optical étendue, which allows it to reach a very high resolution (R=80000) with a very compact instrument ($<<1m^3$). Moreover, its design grants it a high transmission of 40-50\%, which is a significant advantage for characterizing exoplanets, among other objects of interest.

After the laboratory demonstration in the visible, an near-IR version of the spectrometer has been designed\cite{Bourdarot2018}. Following laboratory works on the alignment and calibration procedure of this spectrometer, observing time for an on-sky demonstration of its capabilities has been obtained at Palomar Observatory, and observations occurred on March 14th to 16th, 2022.

This paper first presents in sec.\ref{sec:designspecs} the updated design of the VIPA spectrometer, as well as its specifications as measured in the laboratory. It then discusses in sec.\ref{sec:pipeline} the data reduction pipeline that was developed to extract a spectrum from the raw data. The installation of the spectrometer at the Palomar observatory, and the observations that were conducted are presented in sec.\ref{sec:inobs}. Future developments are discussed in sec.\ref{sec:future}, and a conclusion is given in sec.\ref{sec:conclusion}.

\section{Spectrometer design, specifications, and interfaces}
\label{sec:designspecs}  

We briefly recall in this section the original design\cite{Bourdarot2018} of the spectrometer, and we highlight the small differences with the current one. We also list its specifications, and interfaces, and detail the resolution and transmission of the spectrometer that were measured in the laboratory.

\subsection{Design}
\label{subsec:design}  

A classical échelle spectrograph uses an échelle grating and a cross disperser to project a spectrum over the surface of a detector in two dimensions through a series of orders that each cover a small part of the full bandwidth.

\begin{figure}[t]
\includegraphics[width=16cm]{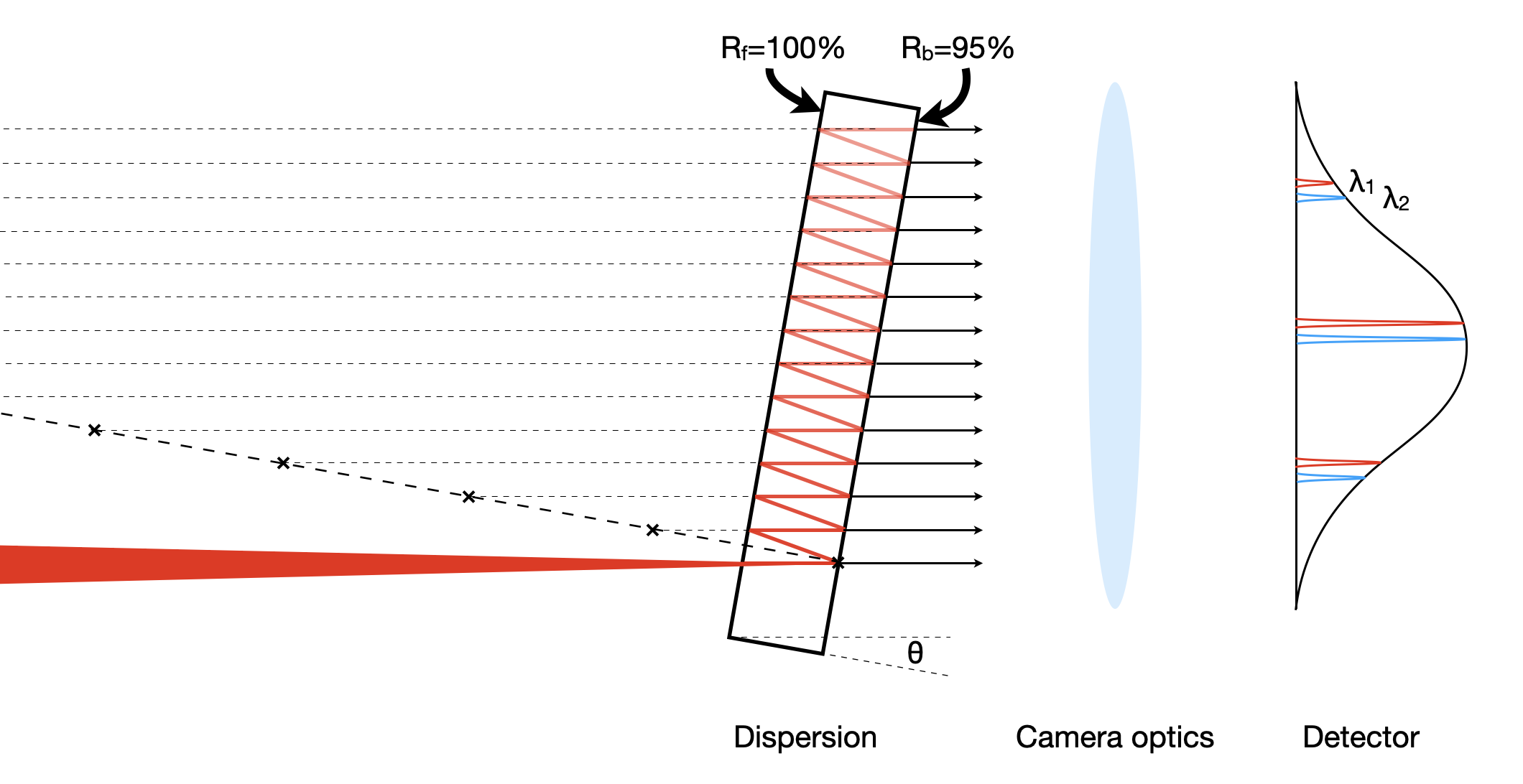}
\centering
\caption{Schematic view of the physics of the VIPA component. A beam is focused at its base, and light reflects between the front surface, which is entirely reflective, and the back surface, which is 95\% reflective. The beams emerging from the VIPA can be seen as if they were coming from reflections on a flat virtual grating, here represented as a dashed black line. These beams, after going through camera optics, induce a series of fringes on a detector, that shift vertically with the wavelength. An Gaussian envelope modulates the intensity of these fringes.}
\label{fig:VIPA}
\end{figure}

In the VIPA spectrometer the échelle grating is replaced with a virtually imaged phased array (VIPA\cite{Shirasaki1996}), which gives it its name. A VIPA is a Fabry-Pérot interferometer used as an angular disperser. The physics of the VIPA is illustrated in fig.\ref{fig:VIPA}. A beam is focused at the entrance of the VIPA, in a thin, 200$\mu m$ slit located at its base. Light bounces back and forth between the front and back surfaces, that have respectively a 100\% and 95\% reflectivity. Each time the beam hits the back surface, a small fraction exits the VIPA. Downstream of the VIPA, light is seen as if it was emerging from an échelle grating, and in this respect, a VIPA can be seen as a virtual échelle grating.

In fig.\ref{fig:VIPA}, a projecting optics is shown to illustrate how the light would be projected on a detector if it was located right after the VIPA. In practice, a cross-disperser is located between the VIPA and the projecting optics, but it is useful to understand the vertical dispersion induced by the VIPA. The same wavelength is projected on the detector at multiple locations, i.e., in several orders. The intensity of each order is modulated by an envelope with a Gaussian profile whose waist $W_{env}$ is set by the waist $W_{0}$ of the incident Gaussian beam, and the ratio of the focal length $f_{cyl}$ of the cylindrical lens that focuses light on the VIPA, and the focal length $f_{cam}$ of the optics that projects light on the detector: $W_{env} = W_{0} \times f_{cam}/f_{cyl}$.

A VIPA has three main advantages over an échelle grating: (1) its efficiency is close to 100\%, (2) it is very small ($\approx$ 1'' sq.), and (3) it is easy to obtain a high dispersion with it. Being very small, it further decreases the size of the instrumentn, and it is easier to manufacture than an échelle grating, making it relatively inexpensive, in spite of a high optical quality ($20nm$ P2V).

Note that a VIPA can only be used with a very small optical étendue. It has a limited, but fairly large, 20\% bandwidth. Perhaps the only drawback of this component is its sensitivity to alignment: light must be focused on a very thin slit ($200\mu m$) located near the base of the VIPA. A misalignment results in a sharp transmission loss, which defeats its key advantage.

\begin{figure}[t]
\includegraphics[width=16cm]{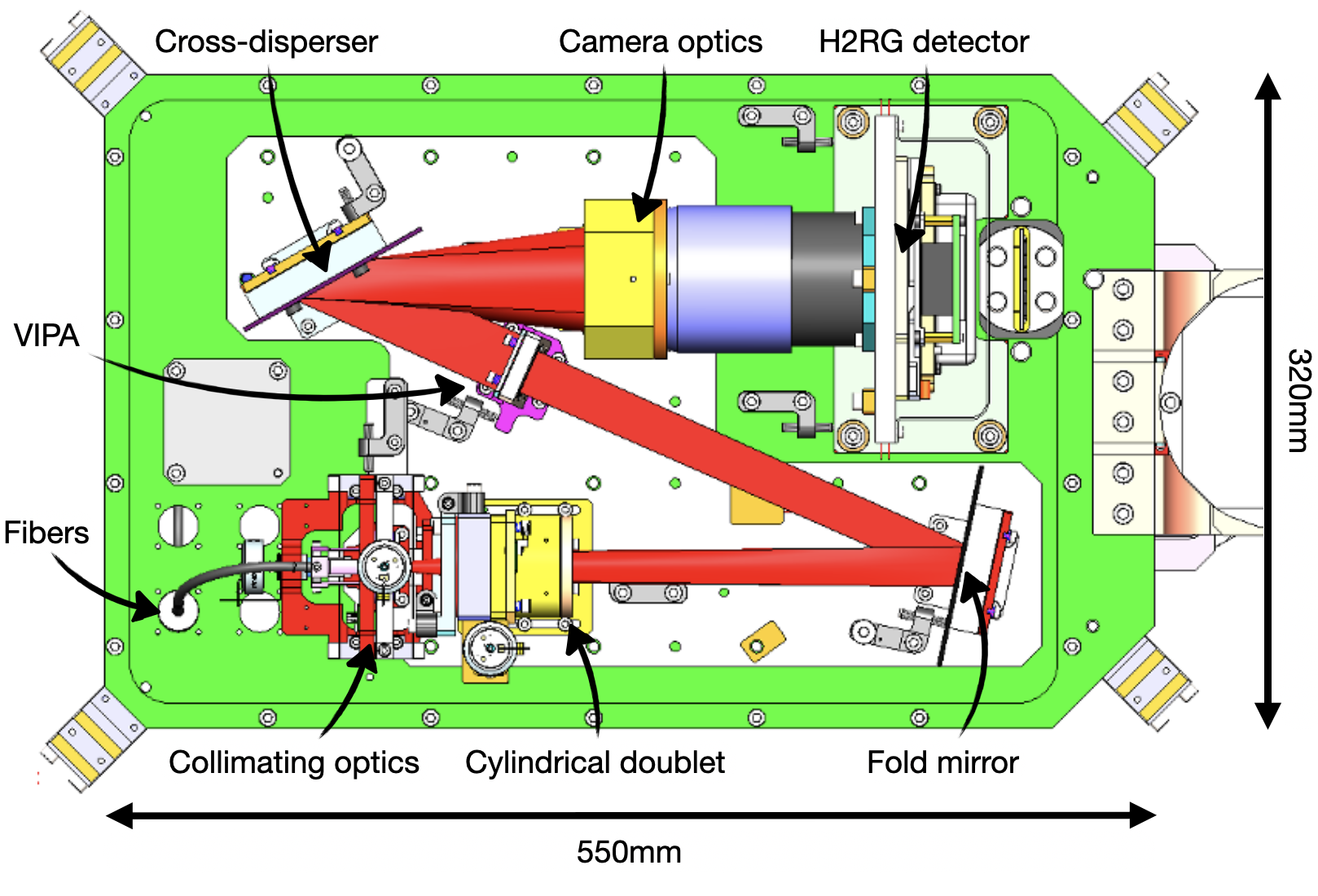}
\centering
\caption{Top-view of the optomechanical design of the VIPA spectrometer, minus the cryostat in which the optics are located. Two fibers carry the light of the targets to be characterized. Light exits the fibers, and is collimated, before being focused with a cylindrical doublet, and folded. The focus is located at the entrance slit of the VIPA, which disperses the light vertically. Light then propagates towards the cross-disperser, and is dispersed horizontally. It is then projected through camera optics onto an H2RG detector.}
\label{fig:VIPA3D}
\end{figure}

The rest of the design is illustrated in fig.\ref{fig:VIPA3D}. Two single-mode fibers can be connected to the spectrometer, so that the light they carry can be analyzed at the same time. Light is transmitted inside a vacuum vessel through fiber feedthroughs, which are connected to two other single-mode fibers inside. The light of these fibers is collimated using a doublet, and focused at the base of the VIPA using a cylindrical lens (and folded with a mirror). Exiting the VIPA, light hits the blazed-grating cross-disperser, before being reimaged onto the surface of an H2RG detector.

Optics are mounted on a 260x400mm breadboard, which is itself mounted on a larger, 320x550mm breadboard that sits about 100mm above the floor of the cryostat. This two-breadboard design was chosen so that the H-band optics could be easily switched with the K-band optics.

The VIPA spectrometer is currently designed to work in the H band. K-band observations are also planned. It uses an H2RG detector, i.e., a low-noise, near-IR detector with 2Kx2K pixels, with the spectrum of a source being projected onto a quarter of the detector (2Kx0.5K). Working at these wavelengths, with this type of detector requires a T=80K cryogenic temperature, which we obtain by placing our spectrometer in a cooled cryostat. 

A primary pump, located next to the instrument, and a secondary pump, attached beneath the cryostat, create a vacuum. A pulse-tube cools down the optics bench inside the cryostat. Its head it attached above the cryostat, and the compressor is located next to the instrument.

Operating at a cryogenic temperature makes the alignment of the optics a bit difficult since the structure deforms with the sharp decrease in temperature. Our initial plan was to use a thermo-mechanical model to predict these deformations, and to apply corrections after aligning the optics at ambient temperature. This proved impractical and we have thus added three cryo-compatible piezo-electric actuators to the original design of the spectrometer

\begin{figure}[t]
\includegraphics[width=16cm]{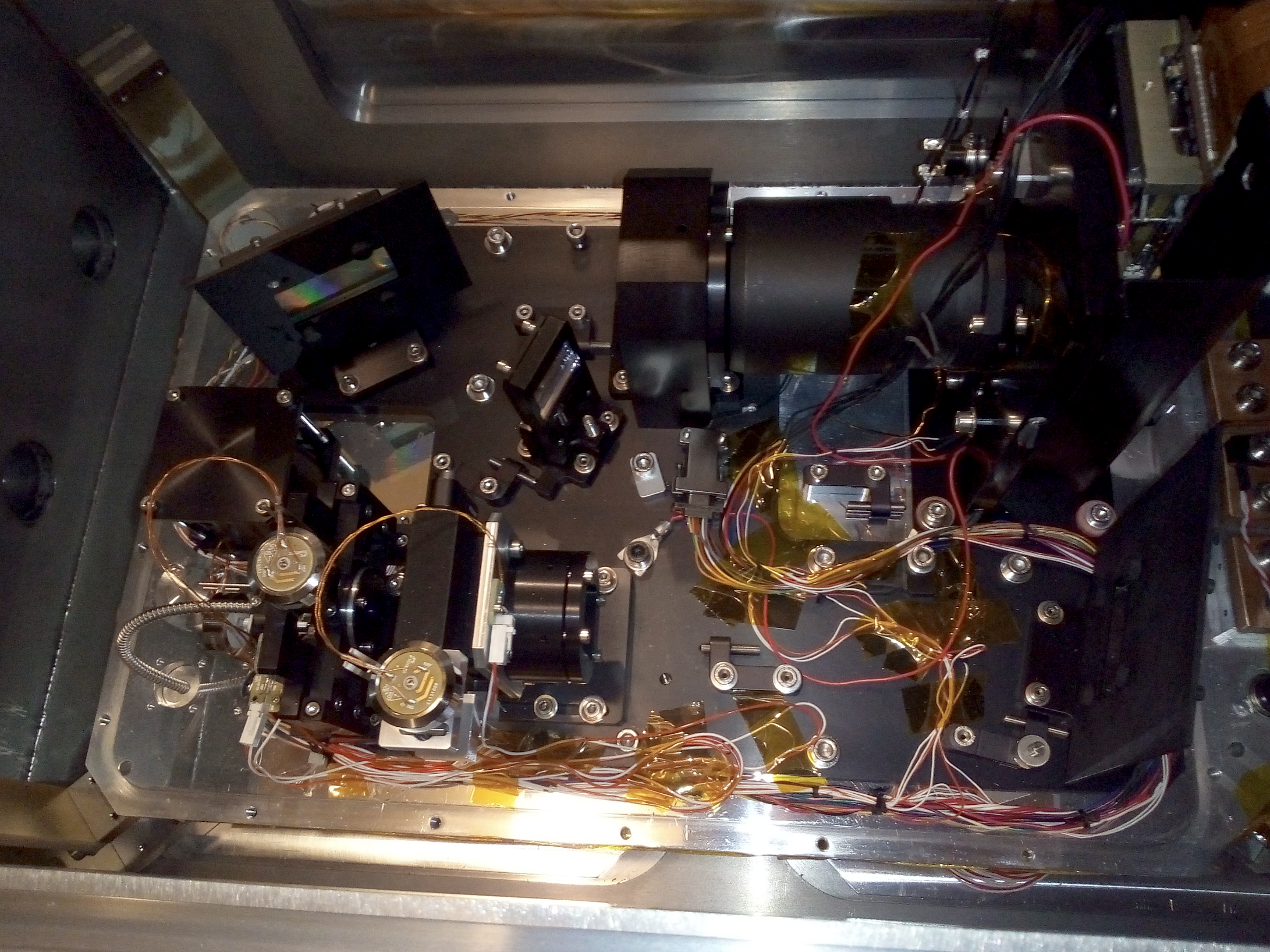}
\centering
\caption{Top-view photography of the optomechanical structure of the VIPA spectrometer. The cryostat that contains it can be seen as well. The orientation of the photography is similar to the drawing in fig.\ref{fig:VIPA3D}. Most of the many wires that can be seen are connected to the temperature sensors, and a few of them are connected to the piezo-electric actuators mounted on the collimating and focusing optics in the bottom left corner.}
\label{fig:VIPA_photo}
\end{figure}

These piezo-electric actuators from JPE are used to adjust three degrees of freedom with a nanometric precision: the height and clocking angle of the cylindrical doublet (and therefore, the height of the beam that hits the VIPA, and the angle between the entrance slit, and the beam), and the distance between the fiber and the collimating doublet. These actuators can be operated both at ambient temperature, and at cryogenic temperature with a vacuum. They can be seen, along with the rest of the optomechanical structure in a photography given in fig.\ref{fig:VIPA_photo}. 

\subsection{Specifications}
\label{subsec:specs}  

In its current design, the VIPA spectrometer observes in a 130nm bandwidth from 1.57$\mu m$ to 1.70$\mu m$. This bandwidth is limited by the specifications of the blazed grating cross-disperser, which were chosen to accommodate the spectra of two sources at the same time on an H2RG detector. When observing two sources, the two sets of orders that are interlaced are separated enough from each other to avoid any cross talk. In practice this separation is overly conservative, given the amount of cross talk, and a third set of orders could be accommodated, or alternatively the bandwidth could be doubled, to around 250nm.

The VIPA spectrometer is designed to achieve a mean resolving power R=80000. It is expected to change along each order, from around R=60000 to R=100000, as well as across orders, from an average resolution of R=75000 to R=85000. The spectrum is sampled through about 80 orders, each covering a 2-3nm range, with a slight overlap between orders, and the same wavelength is sampled once or twice depending on its value.

The total transmission of the spectrometer is expected to be close to $40\%$. It is defined as the ratio between the light that exits the fiber that is connected to the feedthrough mounted on the cryostat, and the one that is detected by the H2RG. A detailed decomposition of the transmission budget is given in \ref{tab:transmission}. It describes the transmission of several sub-elements of the spectrometer, both according to the design, as measured in the laboratory, and expected once some corrections will have been made.



\begin{table}
\begin{center} 
\begin{tabular}{ |c|c|c|c| } 
 \hline
 Component & $T_{Design}$ & $T_{Measured}$ & $T_{Expected}$  \\ \hline 
 Feedthrough & 0.95 & 0.92 (A) | 0.97 (B) & 0.95 \\ 
 Optics & 0.8 & N.A. & 0.69 \\ 
 VIPA component & 0.95 & N.A. & 0.90 \\
 Cross-disperser & 0.7 & 0.41 | 0.79 & 0.79 \\ \hline
 Total spectrometer & 0.5 & 0.21 & 0.47 \\ 
 \hline
\end{tabular}
\caption{\label{tab:transmission} Transmission budget. In addition to the transmission that was assumed from the design, measured transmissions are provided when available. An expected transmission of the spectrometer based on these measurements is given as well. It assumes the replacement of the old grating with the new one, and the removal of the H-band filter.}
\end{center}
\end{table}

A 21\% transmission from the input fiber to the detector plane \emph{without including the detector QE} was measured in the laboratory using a photometer. This value is much lower than the one expected from the design, and we identified three causes for that:
\begin{itemize}
\item the efficiency of the cross-disperser was separately measured to be only 41\%, instead of about 80\% according to the manufacturer. The surface of the cross-disperser appears to have sustained damage after it was initially installed. The cause for this damage has not been identified,
\item the measurement was performed with a wide H-band filter with a mean 90\% transmission that was not part of the original design.
\item the transmission of the feedthrough that was used was measured to be 92\% instead of 95\%. It could be due to some dust, or to a misalignment of the feedthrough's fiber connector. The other feedthrough's transmission, which has been less used, was measured to be 97\%.
\end{itemize}

These causes are not enough to explain the totality of the low, 21\% transmission. Taking into account the three causes described above, the combined transmission of the optics and of the VIPA in this test must have been 62\%, instead of 76\% according to the design. There is no clear reason why the transmission of the VIPA component should be the cause for this difference. A careful alignment of the beam with respect to the VIPA was performed, and such an injection loss in the VIPA appears unlikely. The collimating optics may vignet the beam, however, and may thus explain this difference. This will be investigated in the future, and changes to the collimating optics may be made to avoid this.

A replacement grating was obtained from the manufacturer, but too late for the observations at Palomar Observatory, unfortunately. The transmission of this new grating was measured close to 78\% in broadband, unpolarized light. This value is very close to the one indicated by the manufacturer.

Switching the old grating with the new one, the total transmission of the spectrometer is expected to be 47\%.

This does not include the detector quantum efficiency (QE). A 70\% value was assumed in the original design, but this value corresponds to the minimum QE of an H2RG detector, as provided by its manufacturer.
In practice a much higher value of about 90\% has been measured with the detectors of other instruments such as VLT-SPHERE. 
Such a QE would make the total transmission of the instrument 42\%, which is close to what was expected when designing the instrument.

The H2RG detector that was used for the laboratory tests as well as the on-sky tests at Palomar Observatory is an engineering grade model owned by the University of Montreal. 
The cosmetics of this detector prevent using but a quarter of the detector, and the orders are projected onto the 512x2048 section located next to the lower edge of the H2RG. This is enough to sample the spectrum with at least two pixels per spectral resolution element, satisfying the Shannon-Nyquist criterion.

Even so, a significant fraction of this area is affected by ‘bad pixels’ that return non-usable counts, either because they are systematically too low or too high no matter the signal, or because the variance of the noise is not consistent over time.

\begin{figure}[t]
\includegraphics[width=16cm]{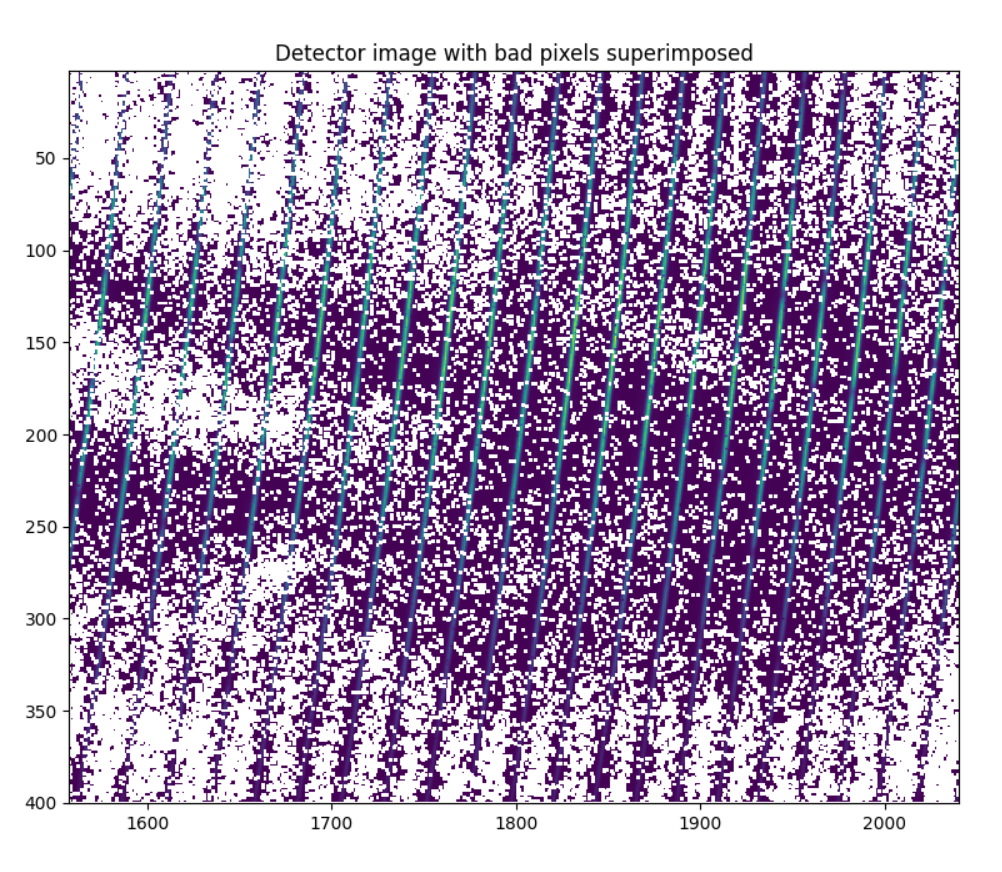}
\centering
\caption{400 pixel square region of the engineering grade detector that was used for the laboratory and on-sky tests. Bad pixels appear in white. The signal is visible too, and is composed of a series of orders that have a 5deg angle with the vertical.}
\label{fig:badpixels}
\end{figure}

Fig.\ref{fig:badpixels} provides a view on a 400 pixels large square region of the detector. It shows both the signal acquired with the VIPA with a broadband source, and the many bad pixels affecting it. This section of the detector is actually one of the less affected.

Another minor issue was encountered with the detector: for some observations, but not all of them, data was lost for 32 columns regularly located on the detector (with a 64 pixels interval). This is likely due to a malfunction of the electronics.

In spite of these limitation, this detector enabled the on-sky tests of the spectrometer, i.e., the demonstration of its capability to provide a spectrum over the desired spectral range, and it allowed for a measurement of its effective resolution, and transmission.

\subsection{Interfaces}
\label{subsec:interfaces}  

\begin{figure}
\includegraphics[width=16cm]{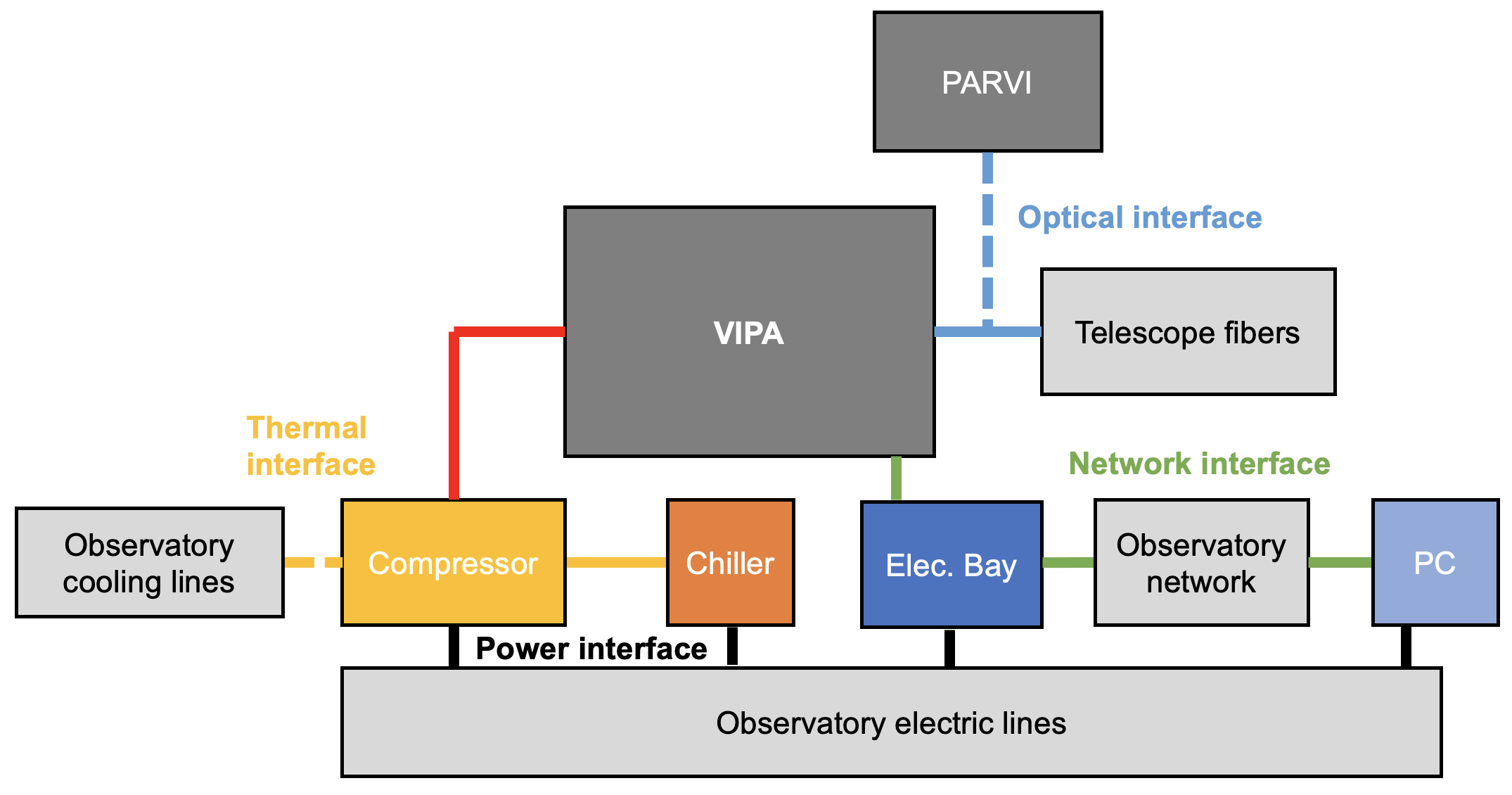}
\centering
\caption{Schematic view of the components of the VIPA spectrometer, and its interfaces with the observatory. Dark grey is used for the VIPA and PARVI instruments, and light gray is used for the observatory.}
\label{fig:interfaces}
\end{figure}

The specifications of the interfaces of the VIPA spectrometer are listed in tab.\ref{tab:interfaces}. The different parts of the spectrometer, and their interfaces with the observatory are illustrated in \ref{fig:interfaces}.

\begin{table}
\begin{center} 
\begin{tabular}{ |c|c|c| } 
 \hline
 Type & Nature & Details \\ \hline 
 Optical & single-mode fiber & FC-APC, SMF28, NA=0.13, 9$\mu m$ core \\
 Mechanical & Volume (L,W,H) & 1x0.8x1$m^3$ \\
 Thermal & Cooling lines & Water or glycol/water mix, T$\ge$27deg C \\
 Electrical & Several high power devices & 15kW total (incl. 5kW pulse-tube compressor) \\
 Network & LAN & IP address (static if possible) \\
 \hline
\end{tabular}
\caption{\label{tab:interfaces} Interfaces of the VIPA spectrometer. The table provides for each type of interface its exact nature, and technical details related to it.}
\end{center}
\end{table}

The cryostat volume is only 0.8x0.5x0.5m$^3$, and it is smaller than the volume indicated in \ref{tab:interfaces}. The latter also takes into account the volume of the secondary pump that is attached beneath the cryostat, as well as legs onto which the cryostat stands.

The numerical aperture (NA) that is assumed for the fibers is $NA=0.13$. The fibers that are used to feed PARVI have the same NA.

The staff at Palomar Observatory greatly helped in the interfacing process. For instance, electric sockets using the European 220V convention were installed specifically for these tests.

\section{Data reduction pipeline}
\label{sec:pipeline}

\begin{figure}[t]
\includegraphics[width=16cm]{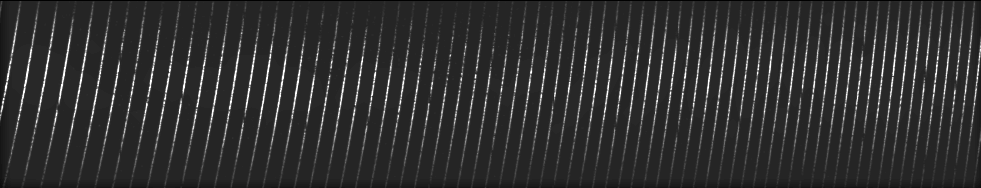}
\centering
\caption{Dark-subtracted order flat image. The dark spots that appear regularly are induced by the electronics read issue discussed in subsec. \ref{subsec:specs}.}
\label{fig:flat}
\end{figure}

\begin{figure}
     \centering
     \begin{subfigure}{0.9\textwidth}
         \centering
         \includegraphics[width=\textwidth]{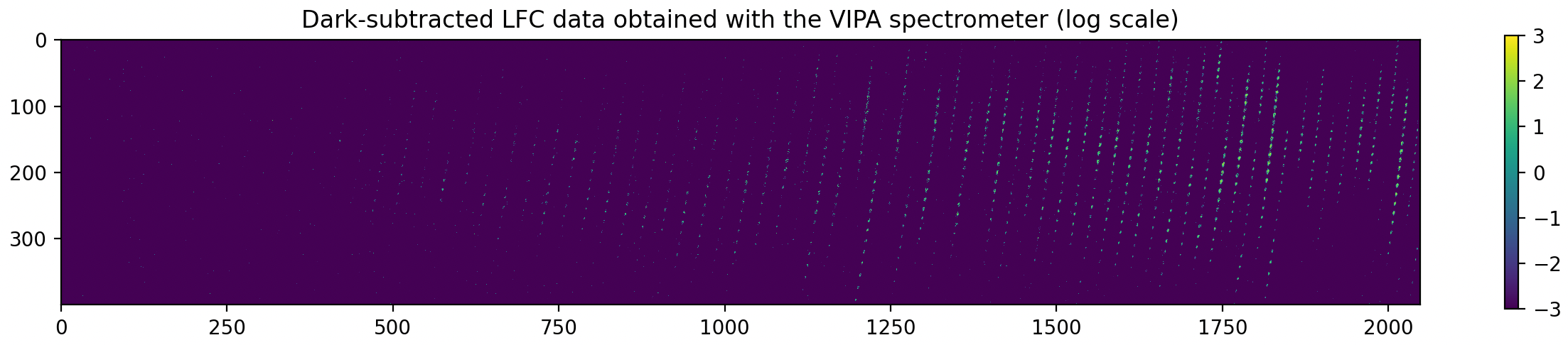}
         \caption{Dark-subtracted order flat image. The dark spots that appear regularly are induced by the electronics read issue discussed in subsec. \ref{subsec:specs}.}
         \label{fig:LFC2D}
     \end{subfigure}
     \\
     \begin{subfigure}{0.9\textwidth}
         \centering
         \includegraphics[width=\textwidth]{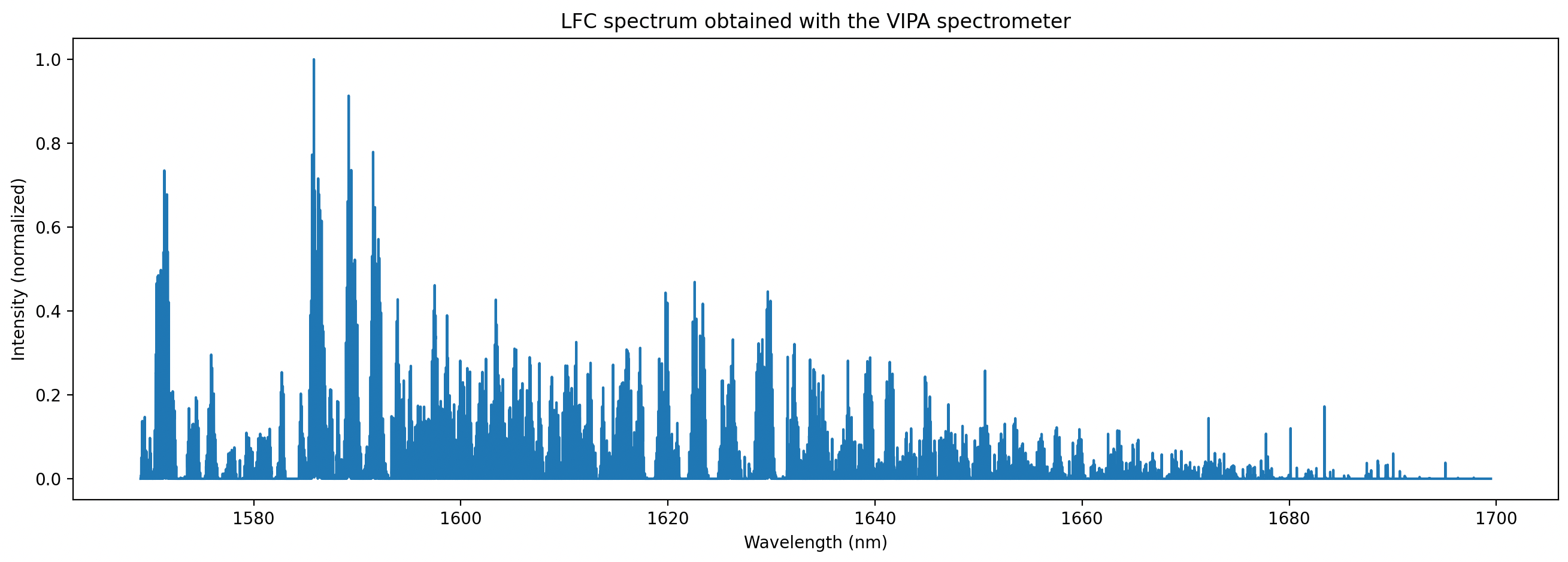}
         \caption{LFC spectrum extracted from the data shown in fig.\ref{fig:LFC2D}}
         \label{fig:LFC1D}
     \end{subfigure}
     \\
     \begin{subfigure}{0.9\textwidth}
         \centering
         \includegraphics[width=\textwidth]{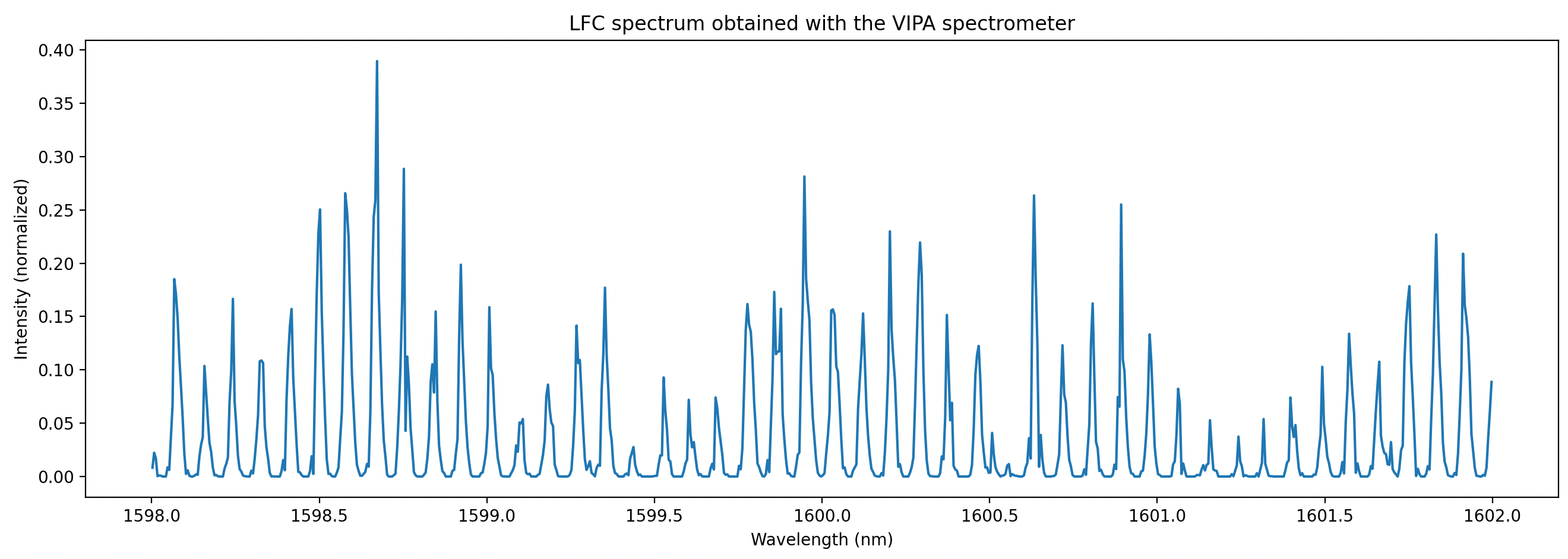}
         \caption{Close-up version of the LFC spectrum shown in fig.\ref{fig:LFC1D}}
         \label{fig:LFC1DZOOM}
     \end{subfigure}
     \caption{LFC spectrum acquired at Palomar observatory with the VIPA spectrometer.}
     \label{fig:LFCdata}
\end{figure}

The reduction of the raw data is similar to that of other échelle spectrographs. After subtracting the dark, the spectrum is extracted from the images by measuring the intensity along the orders. This process uses three prior calibrations:
\begin{itemize}
    \item a dark, which must be acquired in the same condition as the science exposures, and which can be used to identify the bad pixels in the image.
    \item the orders flat, i.e., a spectrum obtained with a broadband, feature-less, and calibrated source, to determine the flux variation along the orders. It was used both to later normalize the spectrum, and to built a quadratic model of the location of the orders in the 2D space of the detector. An order flat is displayed in fig.\ref{fig:flat}.
    \item the pixel map, i.e., a file describing the location of the wavelengths on the detector. It was first obtained using a tunable H-band laser with a picometer precision that we used to scan the system with a 0.1nm resolution, resulting in about 20 points along each order. A quadratic model was built to associate for each order the wavelength with the vertical coordinate of the detector space. The laser frequency comb (LFC) at Palomar observatory was later used to obtain a second pixel map. The LFC samples the J \& H bands at a 10GHz frequency. It has two advantages: (1) it provides this calibration in a single exposure, and (2) it enables a precise monitoring of the stability of a spectrometer. Fig.\ref{fig:LFC2D} shows an example of data acquired with the LFC coupled into the VIPA spectrometer, and fig.\ref{fig:LFC1D} and fig.\ref{fig:LFC1DZOOM} show the extracted spectrum in the full spectral range, and in a 4nm range, respectively.
\end{itemize}

The intensity is estimated using this calibration data at each vertical coordinate (and thus for a specific wavelength) along each order, by fitting a small Gaussian on the data, taking into account the presence of bad pixels. Based on this first processing, a spectrum is derived for a chosen sampling, which is by definition similar or slightly degraded with respect to the resolution of the spectrometer itself.

\begin{figure}
     \centering
     \begin{subfigure}{0.9\textwidth}
         \centering
         \includegraphics[width=\textwidth]{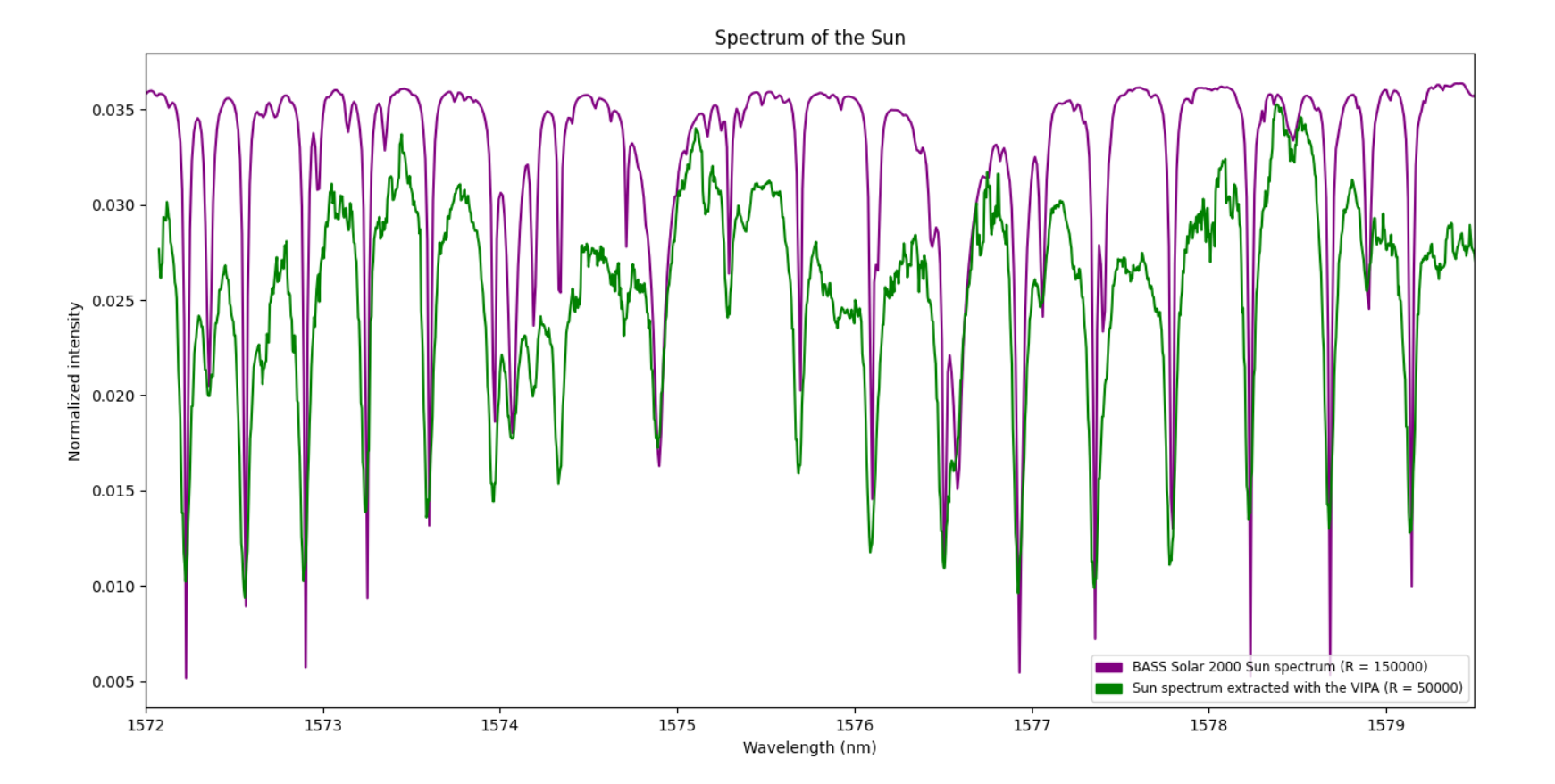}
         \caption{Solar spectrum obtained with the VIPA spectrometer, and compared with the reference solar spectrum from the BASS2000 online database.}
         \label{fig:Sun1}
     \end{subfigure}
     \\
     \begin{subfigure}{0.9\textwidth}
         \centering
         \includegraphics[width=\textwidth]{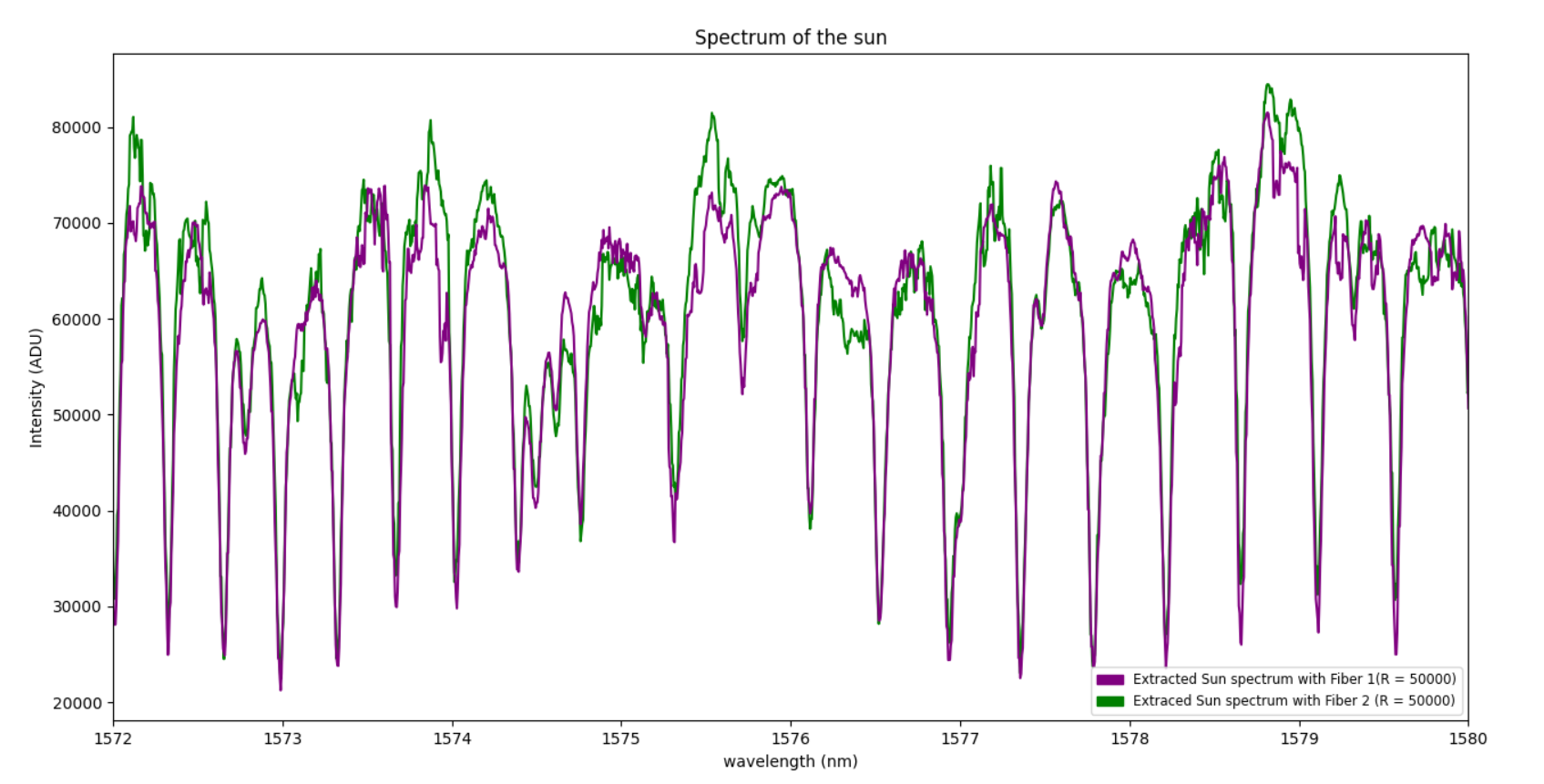}
         \caption{Solar spectra obtained in both channels.}
         \label{fig:Sun2}
     \end{subfigure}
     \caption{Solar spectra obtained with the VIPA spectrometer.}
     \label{fig:Sun}
\end{figure}

Observations of the Sun were conducted from Grenoble. The reduction pipeline was used to extract its spectrum, and the result is shown in fig.\ref{fig:Sun1}, together with a reference IR spectrum obtained from the Bass2000 online database\cite{Delbouille1989}\footnote{https://bass2000.obspm.fr/home.php}. A very good agreement can be observed between the strong absorption features. A low-order modulation of the amplitude of the spectrum obtained with VIPA can be seen as well. It may be related to the atmospheric conditions and composition, in particular a higher water vapour column density at the 200m altitude of Grenoble.

Fig.\ref{fig:Sun2} also shows the spectra obtained with both channels of the VIPA spectrometer, one at a time. Some small differences can be observed, such as the overall higher transmission of channel B with respect to channel A. Other differences may be due to small changes in the atmospheric conditions, and to the cosmetics of the detector, which complicate the data extraction.

\begin{figure}
\includegraphics[width=16cm]{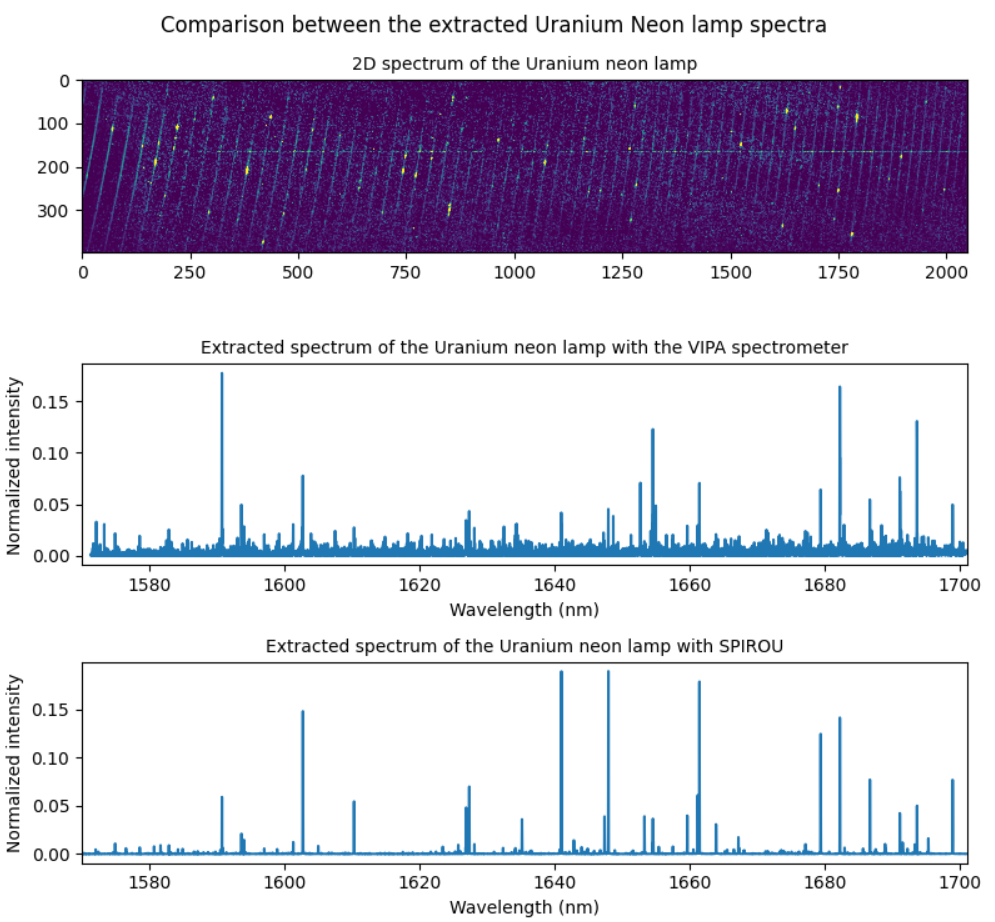}
\centering
\caption{Spectra of Uranium-Neon sources obtained with VIPA and SPIRou. Top: VIPA detector image. Middle: spectrum extracted from the VIPA detector image. Bottom: spectrum extracted for a similar source with SPIRou.}
\label{fig:UR}
\end{figure}

Fig.\ref{fig:UR} shows the spectra of two Uranium-Neon sources obtained with VIPA, and with SPIRou, respectively. The same major emission lines can be observed in both cases. Their intensity is quite different, which can be expected since these measurements were obtained with different sources in slightly different conditions of temperature.

\section{Installation and observations}
\label{sec:inobs} 

Three observation nights with the 5m Hale telescope at Palomar Observatory were allocated for the on-sky test of the VIPA spectrometer on March 14th, 15th, and 16th, 2022.

The PALM300\cite{Meeker2020} adaptive optics system was essential to this demonstration, as well as the fiber injection unit developed for PARVI, and installed in the Cassegrain cage of the telescope.

The instrument and all of the necessary tools to proceed with its operation, alignment and calibration were shipped from the University of Grenoble-Alpes at the end of January.
The H2RG detector was shipped from the University of Montréal in the middle of February.

\subsection{Installation}
\label{subsec:install}

The instrument was installed between February 25, and March 9, 2022. 
A day was necessary to unpack the crates, and move their content to a cleanroom.
It took two days to install the H2RG detector in the cryostat, and perform a first alignment of the optics. This alignment was performed at ambient temperature using a C-RED2 SWIR camera.
Optics were then moved to their cryo position, known from model, and adjusted from previous laboratory tests.

The instrument was then transported to a different room, next to PARVI, and interfaced with the observatory. 
A first cool down was initiated on March 1st, but was stopped following the failure of one of the observatory’s transformer.
The JADE2 card used to communicate with the H2RG detector was unfortunately damaged in the process. Fortunately, a spare card was shipped from the University of Montréal on March 3rd, and received a day later. In the meantime, the observatory staff solved the electrical issue.
A second cool down was initiated on March 4th. The next day, once the detector reached a stable 80K temperature, images were taken using the light of a broadband source connected to the A channel of the instrument.
Light was already going through the VIPA’s entrance slit, and a slight adjustment was made to maximize the instrument throughput. 
Though much less critical, the position of the collimating optics was also slightly adjusted.
New images were then obtained with the broadband source. They revealed that the detector had to be lowered by a few millimeters to avoid an area too rich in bad pixels.
This adjustment cannot be done without opening the instrument. It was performed the next day, following a warm up of the cryostat.
A third cool down of the instrument was then initiated on March 7th.

A flat of the orders was performed on March 8th using the same broadband source that was used for the optics alignment. It was performed for both channels. 
The wavelength solution of the instrument (the pixel map) was derived from a series of images acquired while feeding channels A and B successively with the light of the tunable laser mentioned in \ref{sec:pipeline}.
Another series of images was acquired on the same day using the laser frequency comb (LFC) that is installed next to PARVI. 

\begin{figure}
\includegraphics[width=12cm]{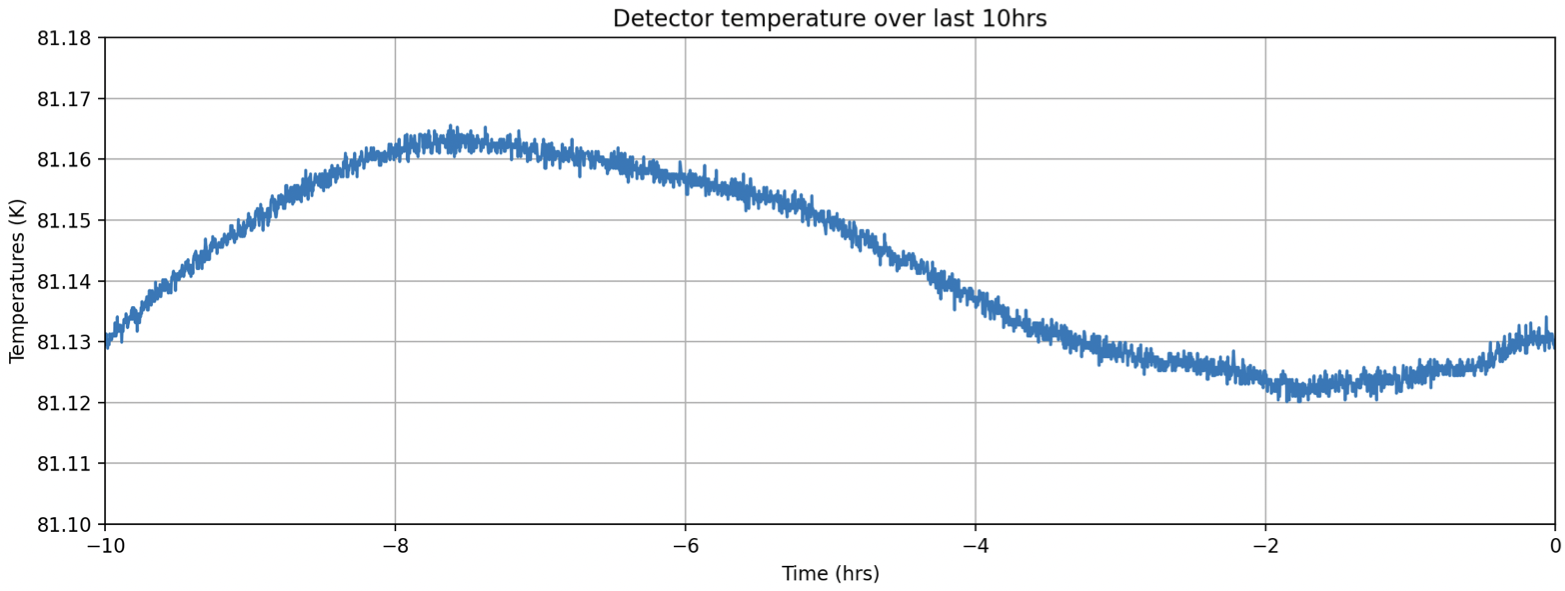}
\centering
\caption{Temperature measured next to the H2RG detector over 10 hours.}
\label{fig:temperature}
\end{figure}

The temperature of the detector was monitored over a few hours to check its stability without a closed-loop temperature control. The result is illustrated in fig.\ref{fig:temperature}. A total variation of 40mK occurs over 10 hours, with a maximum change of 20mK over 1 hour. Note that the VIPA spectrometer was not designed for thermal stability like radial velocity instruments are.

The cryostat was then warmed up, since observations would not occur before a few days. It is our experience that the alignment does not have to be modified after a series of cool down and warm up.

\subsection{Observations}
\label{subsec:obs}

Table \ref{tab:obs} provides the list of the stars that were observed over the allocated three nights, and key observation data.

Note that only stars were observed. Observing exoplanets was not possible because of PARVI's fiber injection unit, whose limiting magnitude is H=11. Even without that, the cosmetics of the detector, the telescope size, and the limited number of nights, would also have prevented this type of observation.

A total of 9 stars were observed, with H magnitudes ranging between 0.8 and 7.1. Brighter stars would have saturated our detector. The faintest stars required a total exposure time of about 1h to reach an SNR close to 100 per spectral resolution element. 

A basic exposure time calculator (ETC) was developed prior to the observations, and was later adjusted to take into account the effective transmission of the telescope ($T_{tel}=0.3$), of the fiber injection unit ($T_{FIU}=0.3$), as well as the Strehl ratio that was obtained that night ($SR\approx0.1$). The latter was limited by a strong dome seeing, and did not change much over the other two nights.

This ETC was successfully used to retrieve the average counts obtained with our spectrometer. This confirmation gives confidence when using this ETC to predict the SNR that could be obtained in other circumstances. For instance, given the same telescope and observing condition, observing an H=8 star would have required about 4h to reach SNR=100, which we deemed was a too long time for our technical tests.

For each star, a first, short observation was performed to check that light was correctly injected in VIPA, i.e., that we could receive a signal, and that the flux was in agreement with the expectation based on the ETC, and on the magnitude of the star. Once we had confirmation that the injection was correct, a longer series of exposures was started.


\begin{table}
\begin{center} 
\begin{tabular}{ |c|c|c|c|c|c| } \hline
Object of interest & H mag & Nramp & NDIT & Reference star & H mag  \\ \hline
V440 Aur & 1.5 & 20 & 64 & HR 2540 & 3.2\\
GJ 388 & 4.8 & 22 & 128 & HR 3975 & 3.5\\
GJ 436 & 6.1 & 30 & 128 & HR 3975 & 4.8\\
Phi Boo & 3.2 & 21 & 64 & HD 143806 & 6.5\\ \hline
HR 3522 & 4.1 & 20 & 64 & HR 3348 & 6.3\\
HD 97813 & 7.1 & 20 & 128 & HD 99966 & 7.5\\
ST Her & 0.8 & 250 & 8 & NA & NA\\ \hline
V440 Aur & 1.5 & 30 & 64 & NA & NA\\
GJ436 & 6.3 & 40 & 128 & HR 4789 & 4.8\\
HR 4789 A & 4.8 & 10 & 128 & NA & NA\\
HR 4789 B & 4.8 & 10 & 128 & NA & NA\\
26 Dra A & 4.2 & 5 & 128 & NA & NA\\
26 Dra B & 6.4 & 15 & 128 & NA & NA \\
\hline
\end{tabular}
\caption{\label{tab:obs} Stars observed during the nights of the 14th, 15th, and 16th of March, 2022. The H magnitude is listed, as well as the number of ramps and NDIT. When reference stars have been observed as well, their name and H magnitude is provided.}
\end{center}
\end{table}

Two of the observed stars (GJ 388, ST Her) are part of the Infrared Telescope Facility (IRTF) spectral library\cite{Rayner2009}. It will thus be possible to compare calibrated, medium resolution (R=2000) spectra, with the spectra acquired with the VIPA spectrometer.

\begin{figure}
\includegraphics[width=12cm]{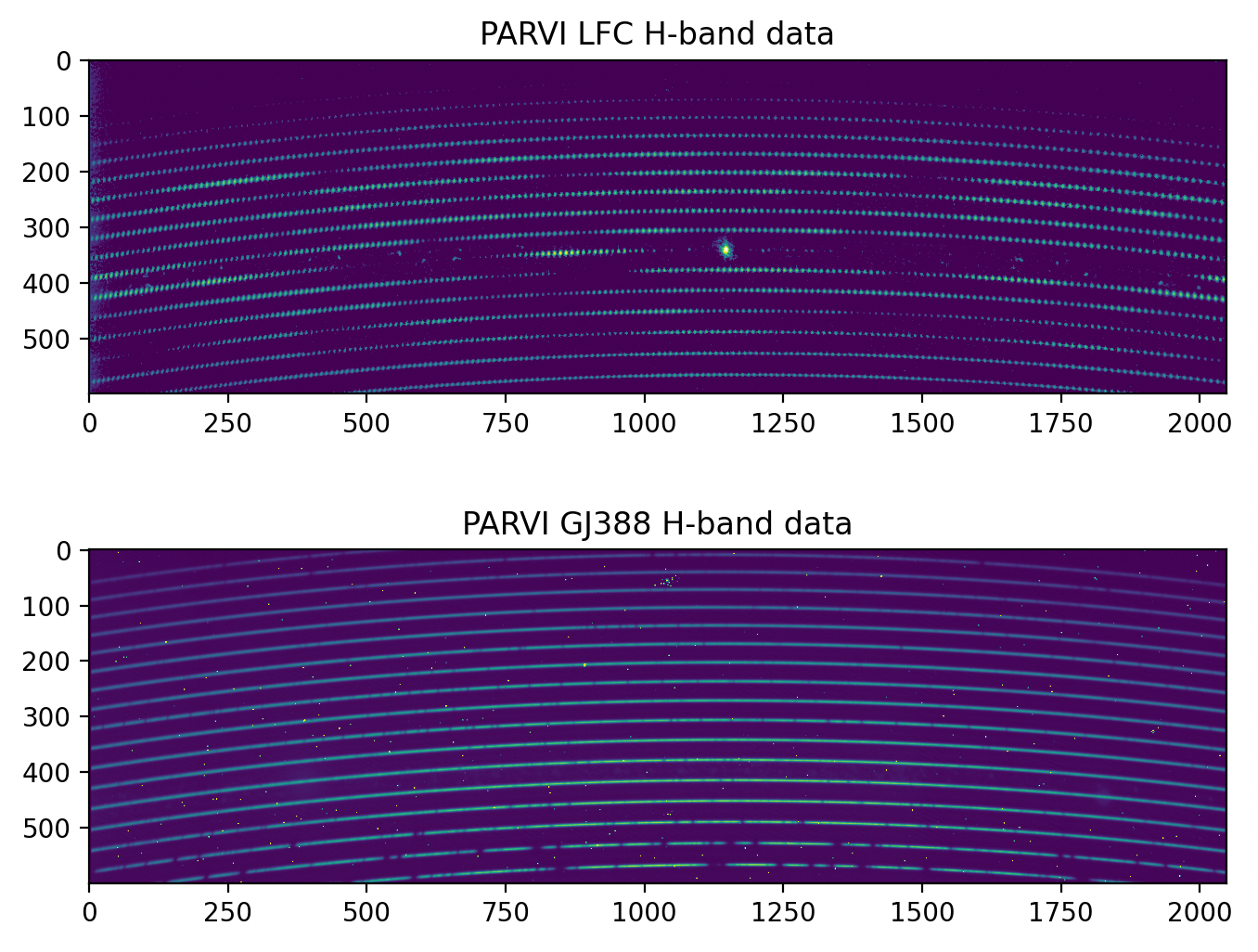}
\centering
\caption{PARVI data: LFC (top, log scale) and GJ388 (bottom). This is an H-band subset of the J+H spectral range in which PARVI observes. The saturated spot in the LFC data is the 1559.9nm pump line.}
\label{fig:PARVI_data}
\end{figure}

GJ 388 was observed both with VIPA alone, and together with PARVI (see fig.\ref{fig:PARVI_data}) using a 50:50 Y-splitter to feed both spectrometers at the same time. This star is also a SPIRou\cite{Donati2018} reference star. The VIPA data for this star will thus be compared to the data obtained with the medium-resolution infrared spectrograph, SpeX\cite{Rayner2003}, along with the PARVI and SPIRou data. PARVI is assumed to have an average transmission of about 20\%.

GJ 436 has a Neptune-like transiting planet with a 2.6d period. It was observed on March 16th, before, during, and after transit. Transmission spectroscopy data may be extracted from this dataset.

HR 4789 is a binary star. The two components were observed one at a time. The observation of HR 4789 B was a bit difficult, and the injection efficiency in this case was likely worse than with HR 4789 A. This is due to the way the star is injected in the single-mode fiber: once the star is set on the fiber, the AO system keeps it at this position in a closed loop. In the case of binary stars that are too close to each other, and/or that present a flux ratio that is too high, the system cannot properly keep the faint companion in, and keeps jumping to the bright companion instead. This is the case with the binary star, for which HR 4789 B is about 2 magnitude fainter than HR 4789 A in the visible, and its angular separation is about 0.4''. A manual offset of the fiber's position was used in order to inject the light of the HR 4789 B, and light was indeed observed on the detector.

26 Dra is another binary star. The separation between the two components of the binary is 0.6'', and the differential H-band magnitude is 2.2. In this case - thanks to the larger separation - the system could automatically keep the fainter component injected in the fiber.

\begin{figure}[t]
\includegraphics[width=12cm]{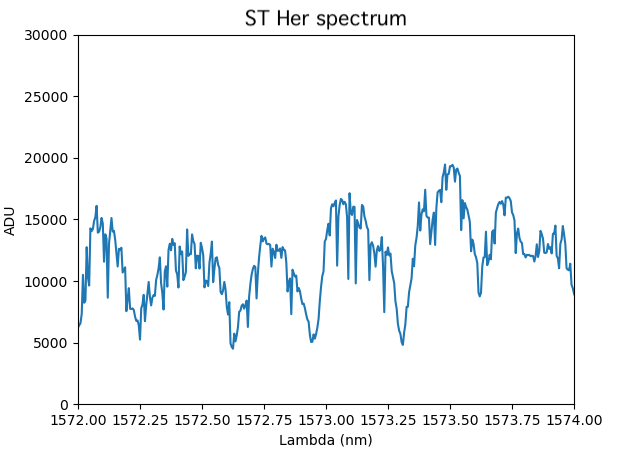}
\centering
\caption{2nm subset of the ST Her spectrum.}
\label{fig:STHer}
\end{figure}

A 2nm large portion of the spectra of ST Her acquired on March 15th is presented in fig.\ref{fig:STHer} as an example. Five broad absorption features can be observed. Many sharp features can also be observed, and are likely artefacts related to the cosmetics of the detector.


\section{Future developments}
\label{sec:future} 

The spectrometer was disassembled immediately after the on-sky tests, and shipped back to IPAG to be modified. Indeed, a number of changes will be made to the instrument, both to improve its current performances, and to prepare for the characterization of exoplanets.

\subsection{Modifications}
\label{subsec:mods} 

A new science grade H2RG detector will replace the engineering grade detector that was used for these tests.
It has already been manufactured, and delivered to IPAG. Its QE was measured by the manufacturer to be 88\% in the near IR. This would make the total transmission of the spectrometer 41\%.
The installation of this new detector in the instrument will occur during fall 2022, and we plan on resuming laboratory tests at the beginning of 2023.

Another change to the instrument will be the replacement of the H-band cross-disperser. 
The new blazed grating will make it possible to cover a twice as large spectral range, i.e., about 250nm, while still allowing for two sources to be characterized at the same time.
The original design assumed a large distance between the orders in order to minimize the difficulty of aligning the two fibers that feed the spectrometer. In practice this distance is larger than required, and we are confident that it can be safely lowered.

While there were good reasons to work in H-band for the first laboratory and on-sky tests (easier alignment process, and compatibility with PARVI’s fiber injection unit), working in the K-band is more interesting from an astrophysics perspective.
Indeed, the K-band concentrates more absorption lines than the H-band, especially of $CH_4$, and $CO$. In addition, young planets such as $\beta$-Pictoris b display a high effective temperature (T=1400K for this planet) such that their detection is more effective in the K-band.
As a matter of fact, the KPIC instrument is specifically designed to characterize planets in K-band.

A K-band set of optics has already been acquired. Mechanical support for these optics will be modified so that piezo-electric actuators can be mounted in the same way they were on the mechanical support of the H-band optics.
Piezo-electric actuators will be modified: absolute encoders will be added. At the moment, without these absolute encoders, and because of hysteresis, it is difficult to close a loop with them when maximizing the throughput of the instrument.
The installation of the optomechanics for the K-band is scheduled for the beginning of 2023.
The K-band calibration of the spectrometer will not be performed using a tunable laser, contrary to what has been done until now.
Instead, fibered gas cells containing either $CH_4$, $CO_2$, or $CO$, will be used to obtain the wavelength solution of the spectrometer. A laser will be used to obtain an additional wavelength reference. The K-band VIPA spectrometer should be ready by summer 2023.

\begin{figure}
\includegraphics[width=12cm]{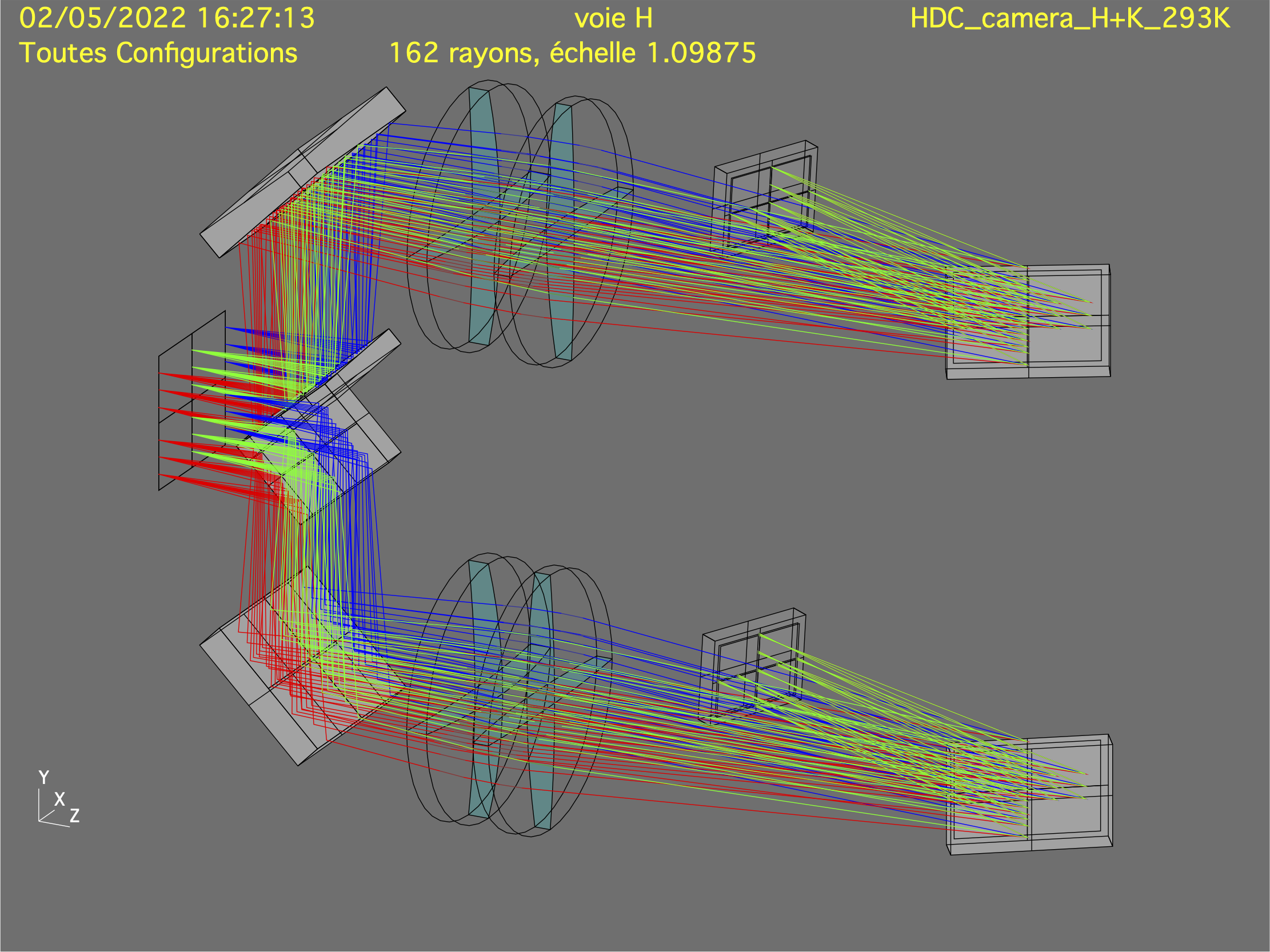}
\centering
\caption{3D view of the optical design of the combined H and K-band mode. Only the dispersive optics (VIPA and cross-disperser) and the camera optics are represented. Note that light is redirected onto the detector using a series of fold mirrors visible on the left side of the image.}
\label{fig:DesignHK}
\end{figure}

Another development is being investigated: the combined characterization of sources in the H and K bands at the same time. A VIPA component may not be manufacturable for such a large band, and the plan would be to use the two sets of optics at the same time.
The current H- and K-band mechanical supports can fit together in the cryostat, albeit with a likely modified upper cover. Fold mirrors would be added to project both spectra onto the detector. A first version of the optical design is presented in Fig.\ref{fig:DesignHK}.

The idea of using photonics components such as arrayed waveguide gratings (AWG) together with VIPA is also a promising one. An AWG can provide a medium to high resolution spectra in an extremely compact format\cite{Gatkine2021}. We will investigate the possibility of using a VIPA/AWG to replace the classical échelle grating/cross-disperser. We will also investigate the possibility of developing a photonics version of a VIPA component.

Finally, the temperature control of the cryostat will be improved to achieve lower temperature variations, in closed-loop, and thus a higher stability. Understanding what limits it may lead to design improvements that could enable an RV mode for this type of spectrometer.

\subsection{Future observations}
\label{subsec:future_obs}

Given its high transmission, the VIPA spectrometer offers a significant advantage over the spectrometers that are used to characterize known exoplanets.

Observations may occur with VLT-SPHERE after the HiRISE fiber injection unit has been added to this instrument. The integration of HiRISE at VLT is scheduled for February 2023. 
The total transmission of HiRISE using with the VIPA spectrometer is expected to be about 8.5\%, instead of 3.6\%.
This gain is due to a twice as high transmission for the spectrometer itself, and the absence of the fiber extraction module (FEM) that is planned to be used with CRIRES+, and that would become unnecessary when using VIPA.
Using VIPA would more than double the transmission of HiRISE when using CRIRES+, thus granting one additional magnitude. 
Note that for the moment HiRISE is expected to only use the H band. 
The H-band collimating optics of the VIPA spectrometer would have to be modified to fit the numerical aperture of the fibers that HiRISE uses, however, otherwise a transmission loss is expected. This change may impact the choice of the new H-band cross-disperser if the fiber core diameter and cladding are different.

Observations may also occur with KPIC. The transmission of NIRSPEC being similar to that of CRIRES+, a similar gain is expected.
Contrary to HiRISE, KPIC is already in operation. 
It only uses the K-band (although observations in the J- and H-bands may become possible).
Connecting KPIC to VIPA could thus happen as early as 2023.

Our ETC predicts that with an ExAO instrument such as VLT-SPHERE (assuming an 85\% Strehl ratio), and with a fiber injection unit with an 85\% transmission (as expected with HiRISE), we could obtain an SNR of 10, per spectral resolution element, in about 5h with an H=18 target. 

This model has two limits. First, HiRISE, as well as KPIC, would be used to look at planets next to a star, and the SNR would thus be also limited by the light of the host star. Second, the spectrum would likely be processed, for instance by cross-correlating it with templates (derived for instance from a BT-SETTL model). Doing so would greatly improve the SNR\cite{Snellen2015}. This technique is being developed at IPAG in the context of another instrument, HARMONI\cite{Thatte2021}, the near-IR, first-light integral field spectrograph of the European ELT.

Observations may also be possible with SCExAO since a fiber injection unit was developed for the high resolution REACH\cite{Kotani2020} H-band spectrometer.

Finally, PAPYRUS\cite{Papyrus2021}, an ExAO system, has recently been installed at the 1.52m telescope at the Haute Provence Observatory in France. It would be much easier to perform future technical tests there. A dedicated fiber injection unit would have to be installed there as well, however.

\section{Conclusion}
\label{sec:conclusion}

The results of the laboratory tests performed at IPAG, and of the on-sky tests performed at the Palomar Observatory demonstrate that the VIPA spectrometer can characterize diffraction-limited sources at an R=80000 mean resolution in a 130nm spectral range in the H-band, with an efficiency close to 40\%. Moreover, its optical design makes it a very compact, and relatively inexpensive instrument whose cost is driven by its detector.

Several hardware changes, including the installation of a dedicated science grade H2RG detector, will be made within one year to enable this spectrometer to be used to characterize planets and other diffraction-limited sources in H and K band. Instruments such as VLT-SPHERE, and KPIC could benefit from using the VIPA spectrometer, and observations my occur as soon as 2023.

\acknowledgments 
This work has been partially supported by the LabEx FOCUS ANR-11-LABX-0013, by a grant from Labex OSUG@2020 (Investissements d’avenir – ANR10 LABX56), and by the European Research Council (ERC) under the European Union's Horizon 2020 research and innovation programme (grant agreement n°866001 - EXACT).

The authors also thank the staff at Palomar Observatory for their efforts in preparing for these tests, and solving out last-minute issues that could have prevented them. 

\bibliography{main} 

\begin{thebibliography}{10}

\bibitem{Beuzit2019}
{Beuzit}, J.~L., {Vigan}, A., {Mouillet}, D., {Dohlen}, K., {Gratton}, R.,
  {Boccaletti}, A., {Sauvage}, J.~F., {Schmid}, H.~M., {Langlois}, M., {Petit},
  C., {Baruffolo}, A., {Feldt}, M., {Milli}, J., {Wahhaj}, Z., {Abe}, L.,
  {Anselmi}, U., {Antichi}, J., {Barette}, R., {Baudrand}, J., {Baudoz}, P.,
  {Bazzon}, A., {Bernardi}, P., {Blanchard}, P., {Brast}, R., {Bruno}, P.,
  {Buey}, T., {Carbillet}, M., {Carle}, M., {Cascone}, E., {Chapron}, F.,
  {Charton}, J., {Chauvin}, G., {Claudi}, R., {Costille}, A., {De Caprio}, V.,
  {de Boer}, J., {Delboulb{\'e}}, A., {Desidera}, S., {Dominik}, C., {Downing},
  M., {Dupuis}, O., {Fabron}, C., {Fantinel}, D., {Farisato}, G., {Feautrier},
  P., {Fedrigo}, E., {Fusco}, T., {Gigan}, P., {Ginski}, C., {Girard}, J.,
  {Giro}, E., {Gisler}, D., {Gluck}, L., {Gry}, C., {Henning}, T., {Hubin}, N.,
  {Hugot}, E., {Incorvaia}, S., {Jaquet}, M., {Kasper}, M., {Lagadec}, E.,
  {Lagrange}, A.~M., {Le Coroller}, H., {Le Mignant}, D., {Le Ruyet}, B.,
  {Lessio}, G., {Lizon}, J.~L., {Llored}, M., {Lundin}, L., {Madec}, F.,
  {Magnard}, Y., {Marteaud}, M., {Martinez}, P., {Maurel}, D., {M{\'e}nard},
  F., {Mesa}, D., {M{\"o}ller-Nilsson}, O., {Moulin}, T., {Moutou}, C.,
  {Orign{\'e}}, A., {Parisot}, J., {Pavlov}, A., {Perret}, D., {Pragt}, J.,
  {Puget}, P., {Rabou}, P., {Ramos}, J., {Reess}, J.~M., {Rigal}, F., {Rochat},
  S., {Roelfsema}, R., {Rousset}, G., {Roux}, A., {Saisse}, M., {Salasnich},
  B., {Santambrogio}, E., {Scuderi}, S., {Segransan}, D., {Sevin}, A.,
  {Siebenmorgen}, R., {Soenke}, C., {Stadler}, E., {Suarez}, M., {Tiph{\`e}ne},
  D., {Turatto}, M., {Udry}, S., {Vakili}, F., {Waters}, L.~B.~F.~M., {Weber},
  L., {Wildi}, F., {Zins}, G., and {Zurlo}, A., ``{SPHERE: the exoplanet imager
  for the Very Large Telescope},'' {\em Astronomy and Astrophysics}~{\bf 631},
  A155 (Nov. 2019).

\bibitem{Macintosh2018}
{Macintosh}, B., {Chilcote}, J.~K., {Bailey}, V.~P., {de Rosa}, R., {Nielsen},
  E., {Norton}, A., {Poyneer}, L., {Wang}, J., {Ruffio}, J.~B., {Graham},
  J.~R., {Marois}, C., {Savransky}, D., and {Veran}, J.-P., ``{The Gemini
  Planet Imager: looking back over five years and forward to the future},'' in
  [{\em Adaptive Optics Systems VI}{\nolinebreak\hspace{0.1em}]},  {Close},
  L.~M., {Schreiber}, L., and {Schmidt}, D., eds., {\em Society of
  Photo-Optical Instrumentation Engineers (SPIE) Conference Series} {\bf
  10703},  107030K (July 2018).

\bibitem{Ahn2021}
{Ahn}, K., {Guyon}, O., {Lozi}, J., {Vievard}, S., {Deo}, V., {Skaf}, N.,
  {Belikov}, R., {Bos}, S.~P., {Bottom}, M., {Currie}, T., {Frazin}, R., {V.
  Gorkom}, K., {Groff}, T.~D., {Haffert}, S.~Y., {Jovanovic}, N., {Kawahara},
  H., {Kotani}, T., {Males}, J.~R., {Martinache}, F., {Mazin}, B., {Miller},
  K., {Norris}, B., {Rodack}, A., and {Wong}, A., ``{SCExAO: a testbed for
  developing high-contrast imaging technologies for ELTs},'' in [{\em Society
  of Photo-Optical Instrumentation Engineers (SPIE) Conference
  Series}{\nolinebreak\hspace{0.1em}]},  {\em Society of Photo-Optical
  Instrumentation Engineers (SPIE) Conference Series} {\bf 11823},  1182303
  (Sept. 2021).

\bibitem{Males2018}
{Males}, J.~R., {Close}, L.~M., {Miller}, K., {Schatz}, L., {Doelman}, D.,
  {Lumbres}, J., {Snik}, F., {Rodack}, A., {Knight}, J., {Van Gorkom}, K.,
  {Long}, J.~D., {Hedglen}, A., {Kautz}, M., {Jovanovic}, N., {Morzinski}, K.,
  {Guyon}, O., {Douglas}, E., {Follette}, K.~B., {Lozi}, J., {Bohlman}, C.,
  {Durney}, O., {Gasho}, V., {Hinz}, P., {Ireland}, M., {Jean}, M., {Keller},
  C., {Kenworthy}, M., {Mazin}, B., {Noenickx}, J., {Alfred}, D., {Perez}, K.,
  {Sanchez}, A., {Sauve}, C., {Weinberger}, A., and {Conrad}, A., ``{MagAO-X:
  project status and first laboratory results},'' in [{\em Adaptive Optics
  Systems VI}{\nolinebreak\hspace{0.1em}]},  {Close}, L.~M., {Schreiber}, L.,
  and {Schmidt}, D., eds., {\em Society of Photo-Optical Instrumentation
  Engineers (SPIE) Conference Series} {\bf 10703},  1070309 (July 2018).

\bibitem{Mawet2018}
{Mawet}, D., {Bond}, C.~Z., {Delorme}, J.~R., {Jovanovic}, N., {Cetre}, S.,
  {Chun}, M., {Echeverri}, D., {Hall}, D., {Lilley}, S., {Wallace}, J.~K., and
  {Wizinowich}, P., ``{Keck Planet Imager and Characterizer: status update},''
  in [{\em Adaptive Optics Systems VI}{\nolinebreak\hspace{0.1em}]},  {Close},
  L.~M., {Schreiber}, L., and {Schmidt}, D., eds., {\em Society of
  Photo-Optical Instrumentation Engineers (SPIE) Conference Series} {\bf
  10703},  1070306 (July 2018).

\bibitem{McLean1998}
{McLean}, I.~S., {Becklin}, E.~E., {Bendiksen}, O., {Brims}, G., {Canfield},
  J., {Figer}, D.~F., {Graham}, J.~R., {Hare}, J., {Lacayanga}, F., {Larkin},
  J.~E., {Larson}, S.~B., {Levenson}, N., {Magnone}, N., {Teplitz}, H., and
  {Wong}, W., ``{Design and development of NIRSPEC: a near-infrared echelle
  spectrograph for the Keck II telescope},'' in [{\em Infrared Astronomical
  Instrumentation}{\nolinebreak\hspace{0.1em}]},  {Fowler}, A.~M., ed., {\em
  Society of Photo-Optical Instrumentation Engineers (SPIE) Conference Series}
  {\bf 3354},  566--578 (Aug. 1998).

\bibitem{Vigan2018}
{Vigan}, A., {Otten}, G.~P.~P.~L., {Muslimov}, E., {Dohlen}, K., {Philipps},
  M.~W., {Seemann}, U., {Beuzit}, J.~L., {Dorn}, R., {Kasper}, M., {Mouillet},
  D., {Baraffe}, I., and {Reiners}, A., ``{Bringing high-spectral resolution to
  VLT/SPHERE with a fiber coupling to VLT/CRIRES+},'' in [{\em Ground-based and
  Airborne Instrumentation for Astronomy VII}{\nolinebreak\hspace{0.1em}]},
  {Evans}, C.~J., {Simard}, L., and {Takami}, H., eds., {\em Society of
  Photo-Optical Instrumentation Engineers (SPIE) Conference Series} {\bf
  10702},  1070236 (July 2018).

\bibitem{Otten2021}
{Otten}, G.~P.~P.~L., {Vigan}, A., {Muslimov}, E., {N'Diaye}, M., {Choquet},
  E., {Seemann}, U., {Dohlen}, K., {Houll{\'e}}, M., {Cristofari}, P.,
  {Phillips}, M.~W., {Charles}, Y., {Baraffe}, I., {Beuzit}, J.~L., {Costille},
  A., {Dorn}, R., {El Morsy}, M., {Kasper}, M., {Lopez}, M., {Mordasini}, C.,
  {Pourcelot}, R., {Reiners}, A., and {Sauvage}, J.~F., ``{Direct
  characterization of young giant exoplanets at high spectral resolution by
  coupling SPHERE and CRIRES+},'' {\em Astronomy and Astrophysics}~{\bf 646},
  A150 (Feb. 2021).

\bibitem{Brucalassi2018SPIE}
{Brucalassi}, A., {Dorn}, R.~J., {Follert}, R., {Hatzes}, A., {Bristow}, P.,
  {Seemann}, U., {Cumani}, C., {Eschbaumer}, S., {Haimerl}, A., {Haug}, M.,
  {Heiter}, U., {Hinterschuster}, R., {Ives}, D.~J., {Jung}, Y., {Kerber}, F.,
  {Klein}, B., {Lavail}, A., {Lizon}, J.~L., {Marquart}, T., {Moins}, C.,
  {Molina-Conde}, I., {Oliva}, E., {Pasquini}, L., {Paufique}, J., {Piskunov},
  N., {Stegmeier}, J., {Stempels}, E., {Tordo}, S., {Valenti}, E.,
  {Anwand-Heerwart}, H., {Hauptner}, K., {Jeep}, P., {Marvin}, C., {Reiners},
  A., {Rhode}, P., {Schmidt}, C., and {Umlauf}, T., ``{Full system test and
  early preliminary acceptance Europe results for CRIRES+},'' in [{\em
  Ground-based and Airborne Instrumentation for Astronomy
  VII}{\nolinebreak\hspace{0.1em}]},  {Evans}, C.~J., {Simard}, L., and
  {Takami}, H., eds., {\em Society of Photo-Optical Instrumentation Engineers
  (SPIE) Conference Series} {\bf 10702},  1070239 (July 2018).

\bibitem{Kotani2020}
{Kotani}, T., {Kawahara}, H., {Ishizuka}, M., {Jovanovic}, N., {Vievard}, S.,
  {Lozi}, J., {Sahoo}, A., {Guyon}, O., {Yoneta}, K., and {Tamura}, M.,
  ``{Extremely high-contrast, high spectral resolution spectrometer REACH for
  the Subaru Telescope},'' in [{\em Society of Photo-Optical Instrumentation
  Engineers (SPIE) Conference Series}{\nolinebreak\hspace{0.1em}]},  {\em
  Society of Photo-Optical Instrumentation Engineers (SPIE) Conference Series}
  {\bf 11448},  1144878 (Dec. 2020).

\bibitem{Kotani2018}
{Kotani}, T., {Tamura}, M., {Nishikawa}, J., {Ueda}, A., {Kuzuhara}, M.,
  {Omiya}, M., {Hashimoto}, J., {Ishizuka}, M., {Hirano}, T., {Suto}, H.,
  {Kurokawa}, T., {Kokubo}, T., {Mori}, T., {Tanaka}, Y., {Kashiwagi}, K.,
  {Konishi}, M., {Kudo}, T., {Sato}, B., {Jacobson}, S., {Hodapp}, K.~W.,
  {Hall}, D.~B., {Aoki}, W., {Usuda}, T., {Nishiyama}, S., {Nakajima}, T.,
  {Ikeda}, Y., {Yamamuro}, T., {Morino}, J.-I., {Baba}, H., {Hosokawa}, K.,
  {Ishikawa}, H., {Narita}, N., {Kokubo}, E., {Hayano}, Y., {Izumiura}, H.,
  {Kambe}, E., {Kusakabe}, N., {Kwon}, J., {Ikoma}, M., {Hori}, Y., {Genda},
  H., {Fukui}, A., {Fujii}, Y., {Kawahara}, H., {Olivier}, G., {Jovanovic}, N.,
  {Harakawa}, H., {Hayashi}, M., {Hidai}, M., {Machida}, M., {Matsuo}, T.,
  {Nagata}, T., {Ogihara}, M., {Takami}, H., {Takato}, N., {Terada}, H., and
  {Oh}, D., ``{The infrared Doppler (IRD) instrument for the Subaru telescope:
  instrument description and commissioning results},'' in [{\em Ground-based
  and Airborne Instrumentation for Astronomy VII}{\nolinebreak\hspace{0.1em}]},
   {Evans}, C.~J., {Simard}, L., and {Takami}, H., eds., {\em Society of
  Photo-Optical Instrumentation Engineers (SPIE) Conference Series} {\bf
  10702},  1070211 (July 2018).

\bibitem{Crass2019}
{Crass}, J., {Bechter}, A., {Bechter}, E., {Beichman}, C., {Blake}, C.,
  {Crepp}, J.~R., {Coutts}, D., {Feger}, T., {Halverson}, S., {Harris}, R.~J.,
  {Jovanovic}, N., {Mawet}, D., {Plavchan}, P., {Schwab}, C., {Vasisht}, G.,
  {Wallace}, J.~K., and {Wang}, J., ``{The need for single-mode fiber-fed
  spectrographs},'' in [{\em Bulletin of the American Astronomical
  Society}{\nolinebreak\hspace{0.1em}]},   {\bf 51},  122 (Sept. 2019).

\bibitem{Bourdarot2017}
{Bourdarot}, G., {Le Coarer}, E., {Bonfils}, X., {Alecian}, E., {Rabou}, P.,
  and {Magnard}, Y., ``{NanoVipa: a miniaturized high-resolution echelle
  spectrometer, for the monitoring of young stars from a 6U Cubesat},'' {\em
  CEAS Space Journal}~{\bf 9},  411--419 (Dec. 2017).

\bibitem{Bourdarot2018}
{Bourdarot}, G., {Le Coarer}, E., {Mouillet}, D., {Correia}, J.-J., {Jocou},
  L., {Rabou}, P., {Carlotti}, A., {Bonfils}, X., {Artigau}, E., {Vallee}, P.,
  {Doyon}, R., {Forveille}, T., {Stadler}, E., {Magnard}, Y., and {Vigan}, A.,
  ``{Experimental test of a 40 cm-long R=100 000 spectrometer for exoplanet
  characterisation},'' in [{\em Ground-based and Airborne Instrumentation for
  Astronomy VII}{\nolinebreak\hspace{0.1em}]},  {Evans}, C.~J., {Simard}, L.,
  and {Takami}, H., eds., {\em Society of Photo-Optical Instrumentation
  Engineers (SPIE) Conference Series} {\bf 10702},  107025Y (July 2018).

\bibitem{Shirasaki1996}
{Shirasaki}, M., ``{Large angular dispersion by a virtually imaged phased array
  and its application to a wavelength demultiplexer},'' {\em Optics
  Letters}~{\bf 21},  366--368 (Mar. 1996).

\bibitem{Delbouille1989}
{Delbouille}, L., {Roland}, G., and {Neven}, L.,  [{\em {Atlas photometrique du
  spectre solaire de [lambda] 3000 a [lambda] 10000.
  Vol.2}}{\nolinebreak\hspace{0.1em}]} (1989).

\bibitem{Meeker2020}
{Meeker}, S.~R., {Truong}, T.~N., {Roberts}, J.~E., {Shelton}, J.~C.,
  {Fregoso}, S.~F., {Burruss}, R.~S., {Dekany}, R.~G., {Wallace}, J.~K.,
  {Baker}, J.~W., {Heffner}, C.~M., {Mawet}, D., {Rykoski}, K.~M., {Tesch},
  J.~A., and {Vasisht}, G., ``{Design and performance of the PALM-3000 3.5 kHz
  upgrade},'' in [{\em Society of Photo-Optical Instrumentation Engineers
  (SPIE) Conference Series}{\nolinebreak\hspace{0.1em}]},  {\em Society of
  Photo-Optical Instrumentation Engineers (SPIE) Conference Series} {\bf
  11448},  114480W (Dec. 2020).

\bibitem{Rayner2009}
{Rayner}, J.~T., {Cushing}, M.~C., and {Vacca}, W.~D., ``{The Infrared
  Telescope Facility (IRTF) Spectral Library: Cool Stars},'' {\em Astrophysical
  Journal Supplement Series}~{\bf 185},  289--432 (Dec. 2009).

\bibitem{Donati2018}
{Donati}, J.-F., {Kouach}, D., {Lacombe}, M., {Baratchart}, S., {Doyon}, R.,
  {Delfosse}, X., {Artigau}, {\'E}., {Moutou}, C., {H{\'e}brard}, G., {Bouchy},
  F., {Bouvier}, J., {Alencar}, S., {Saddlemyer}, L., {Par{\`e}s}, L., {Rabou},
  P., {Micheau}, Y., {Dolon}, F., {Barrick}, G., {Hernandez}, O., {Wang},
  S.~Y., {Reshetov}, V., {Striebig}, N., {Challita}, Z., {Carmona}, A.,
  {Tibault}, S., {Martioli}, E., {Figueira}, P., {Boisse}, I., and {Pepe}, F.,
  ``{SPIRou: A NIR Spectropolarimeter/High-Precision Velocimeter for the
  CFHT},'' in [{\em Handbook of Exoplanets}{\nolinebreak\hspace{0.1em}]},
  {Deeg}, H.~J. and {Belmonte}, J.~A., eds.,  107 (2018).

\bibitem{Rayner2003}
{Rayner}, J.~T., {Toomey}, D.~W., {Onaka}, P.~M., {Denault}, A.~J.,
  {Stahlberger}, W.~E., {Vacca}, W.~D., {Cushing}, M.~C., and {Wang}, S.,
  ``{SpeX: A Medium-Resolution 0.8-5.5 Micron Spectrograph and Imager for the
  NASA Infrared Telescope Facility},'' {\em Publications of the Astronomical
  Society of the Pacific}~{\bf 115},  362--382 (Mar. 2003).

\bibitem{Gatkine2021}
{Gatkine}, P., {Jovanovic}, N., {Jewell}, J., {Wallace}, J.~K., and {Mawet},
  D., ``{An on-chip astrophotonic spectrograph with a resolving power of
  12,000},'' in [{\em UV/Optical/IR Space Telescopes and Instruments:
  Innovative Technologies and Concepts X}{\nolinebreak\hspace{0.1em}]},  {\em
  Society of Photo-Optical Instrumentation Engineers (SPIE) Conference Series}
  {\bf 11819},  118190I (Aug. 2021).

\bibitem{Snellen2015}
{Snellen}, I., {de Kok}, R., {Birkby}, J.~L., {Brandl}, B., {Brogi}, M.,
  {Keller}, C., {Kenworthy}, M., {Schwarz}, H., and {Stuik}, R., ``{Combining
  high-dispersion spectroscopy with high contrast imaging: Probing rocky
  planets around our nearest neighbors},'' {\em Astronomy and
  Astrophysics}~{\bf 576},  A59 (Apr. 2015).

\bibitem{Thatte2021}
{Thatte}, N., {Tecza}, M., {Schnetler}, H., {Neichel}, B., {Melotte}, D.,
  {Fusco}, T., {Ferraro-Wood}, V., {Clarke}, F., {Bryson}, I., {O'Brien}, K.,
  {Mateo}, M., {Garcia Lorenzo}, B., {Evans}, C., {Bouch{\'e}}, N., {Arribas},
  S., and {HARMONI Consortium}, ``{HARMONI: the ELT's First-Light Near-infrared
  and Visible Integral Field Spectrograph},'' {\em The Messenger}~{\bf 182},
  7--12 (Mar. 2021).

\bibitem{Papyrus2021}
{Muslimov}, E., {Levraud}, N., {Chambouleyron}, V., {Boudjema}, I., {Lau}, A.,
  {Caillat}, A., {Pedreros}, F., {Otten}, G., {El Hadi}, K., {Joaquina}, K.,
  {Maxime}, M., {El Morsy}, M., {Beltramo-Martin}, O., {F{\'e}tick}, R., {Ke},
  Z., {Sauvage}, J.~F., {Neichel}, B., {Fusco}, T., {Schmitt}, J., {Le Van
  Suu}, A., {Charton}, J., {Schimpf}, A., {Martin}, B., {Dintrono}, F.,
  {Esposito}, S., and {Pina}, E., ``{Current status of PAPYRUS: the pyramid
  based adaptive optics system at LAM/OHP},'' in [{\em Society of Photo-Optical
  Instrumentation Engineers (SPIE) Conference
  Series}{\nolinebreak\hspace{0.1em}]},  {\em Society of Photo-Optical
  Instrumentation Engineers (SPIE) Conference Series} {\bf 11876},  118760H
  (Sept. 2021).

\end{thebibliography}
\bibliographystyle{spiebib} 

\end{document}